\documentclass[singlespacing, 11pt]{article}

\usepackage{lmodern}
\usepackage{amssymb, amsmath}
\usepackage{ifxetex,ifluatex}
\usepackage[T1]{fontenc}
\usepackage[utf8]{inputenc}

\IfFileExists{microtype.sty}{%
\usepackage{microtype}
\UseMicrotypeSet[protrusion]{basicmath}  % disable protrusion for tt fonts
}{}

\usepackage{color}
\usepackage{xcolor}
\usepackage[margin=1in]{geometry}
\usepackage{pdflscape}

\usepackage{calc}
\usepackage{titlesec}

\titleformat*{\section}{\fontfamily{lmss}\large\bfseries}{}{}{}
\titleformat*{\subsection}{\fontfamily{lmss}\large\itshape}{}{}{}
\titleformat{\subsubsection}[runin]{\fontfamily{lmss}\normalsize\itshape}{}{0pt}{}

\titlespacing*{\section}      {0pt}{3.5ex  plus 1ex minus .2ex}{1.5ex plus .2ex}
\titlespacing*{\subsection}   {0pt}{3.25ex plus 1ex minus .2ex}{1.0ex plus .1ex}
\usepackage[originalparameters]{ragged2e}  % prevent hyphentation

\renewcommand{\raggedright}{\RaggedRight}

\usepackage{mdwlist}

\usepackage{etoolbox}
\usepackage{setspace}
\AtBeginEnvironment{quote}{\singlespacing}
\AtBeginEnvironment{tabular}{\singlespacing}
\AtBeginEnvironment{longtable}{\singlespacing}
\newlength{\normparindent}
\newlength{\normleftskip}
\newlength{\normparskip}
\usepackage{ctable}
\usepackage{booktabs}
\usepackage{longtable}
\usepackage{multirow}

\usepackage{array}

\usepackage{afterpage}

\belowbottomsep=\belowrulesep

\usepackage[export]{adjustbox}  % sizing
\usepackage[section]{placeins}  % provides \FloatBarrier
\usepackage{graphicx}

\makeatletter
\def\maxwidth{\ifdim\Gin@nat@width>\linewidth\linewidth\else\Gin@nat@width\fi}
\def\maxheight{\ifdim\Gin@nat@height>\textheight\textheight\else\Gin@nat@height\fi}
\makeatother
\setkeys{Gin}{width=\maxwidth,height=\maxheight,keepaspectratio}

\usepackage[width=\linewidth,justification=centering]{caption}
\usepackage[width=\linewidth,singlelinecheck=off,justification=centering]{subcaption}

\DeclareGraphicsExtensions{.pdf,.png}

\usepackage{footnote}   % load xcolor first
\makesavenoteenv{tabular}
\makesavenoteenv{table}

\usepackage[flushmargin, bottom]{footmisc}  % http://www.compsoc.man.ac.uk/~moz/www-user/help/footnotes.php

\usepackage{endnotes}

\patchcmd{\enoteformat}{1.8em}{0pt}{}{}

\usepackage[authoryear,longnamesfirst,sort]{natbib}
\setcitestyle{authoryear,aysep={},yysep={,},notesep={:}}

\usepackage[unicode=true]{hyperref}
\hypersetup{breaklinks=true,
            pdfauthor={Elizabeth Roberto},
            pdftitle={The Divergence Index: A Decomposable Measure of Segregation and Inequality},
            pdfkeywords={segregation, inequality, diversity, divergence, entropy, decomposition},
            colorlinks=true,
            citecolor=black,
            urlcolor=blue,
            linkcolor=black,
            pdfborder={0 0 0},
            pdfcreator={LaTeX via pandoc}}
\title{{\singlespacing The Divergence Index: \\ A Decomposable Measure of Segregation and Inequality}}

\author{Elizabeth Roberto\thanks{Direct correspondence to Elizabeth Roberto, Rice University, Department of Sociology, 6100 Main Street, MS-28, Houston, TX 77005. Phone: +1 713-348-3466. Email: eroberto@rice.edu.} \thanks{Thank you to Richard Breen, Scott Page, Russell Golman, Peter Rich, Jackelyn Hwang, Jim Elliott, and Jeremy Fiel for their valuable feedback on this research.  This research was supported in part by the James S. McDonnell Foundation Postdoctoral Fellowship Award in Studying Complex Systems.}}

\date{February 20, 2024}

\begin{document}
\setlength{\parskip}{0pt plus 0pt}
\maketitle
\setlength{\parskip}{\normparskip}

\addcontentsline{toc}{section}{Abstract}
\begin{singlespace}
\begin{abstract}
\normalsize 
Decomposition analysis is a critical tool for understanding the social and spatial dimensions of segregation and diversity. In this paper, I highlight the conceptual, mathematical, and empirical distinctions between segregation and diversity and introduce the Divergence Index as a decomposable measure of segregation. Scholars have turned to the Information Theory Index as the best alternative to the Dissimilarity Index in decomposition studies, however it measures diversity rather than segregation. I demonstrate the importance of preserving this conceptual distinction with a decomposition analysis of  segregation and diversity in U.S. metropolitan areas from 1990 to 2010, which shows that the Information Theory Index has tended to decrease, particularly within cities, while the Divergence Index has tended to increase, particularly within suburbs.  Rather than being a substitute for measures of diversity, the Divergence Index complements existing measures by enabling the analysis and decomposition of segregation alongside diversity.
\end{abstract}
\end{singlespace}

\begin{center}
\end{center}

\newpage

\let\origfootnote\footnote
\let\origfootnotemark\footnotemark
\let\origfootnotetext\footnotetext

\let\footnote=\endnote
\renewcommand{\notesname}{}
\renewcommand{\footnotemark}{\endnotemark}
\renewcommand{\footnotetext}{\endnotetext}

\hypertarget{introduction}{%
\section{Introduction}\label{introduction}}

Decomposable measures of segregation and diversity are essential tools
for understanding the multi-level dynamics of social and spatial
phenomena. For example, recent studies have shown that the U.S.
population is becoming more racially and ethnically diverse as a whole,
but also that the speed and extent of this change varies greatly across
different geographic scales
\citep{Hall:2016wl, Frey:2015uw, Lichter:2015gz}, as rising residential
segregation among municipalities offsets increasing opportunities for
inter-racial and -ethnic contact at the neighborhood level
\citep{Lichter:2015gz}. These dynamics have important implications not
only for the social integration of everyday life but also for the
spatial construction of social divisions that continue to segregate
housing and job opportunities, school districts, gerrymandered voting
jurisdictions, and other socially consequential spaces.

Because of the importance and multiscalar complexity of these spatial
dynamics, it is important to develop and utilize new decomposable
measures that extend beyond more conventional measures to allow
decomposition both within and between different groups and categorical
(often spatial) units. In the present study, I introduce and demonstrate
such a measure, which I call the Divergence Index.

The Dissimilarity Index
\citep{Jahn:1947vj, Duncan:1955ve, Taeuber:1965us} is the most widely
used measure of residential segregation, but it is not additively
decomposable into the segregation occurring within and between groups or
places \citep{Reardon:2002vt, Reardon:2004vl, Theil:1972vb}.
\label{R1_1d_p1} Due to this limitation, scholars have turned to the
Information Theory Index
\citep{Reardon:2002vt, Reardon:2004vl, Theil:1971jx, White:1986vx} to
conduct decomposition analyses of segregation
\citep{Bischoff:2008bw, Farrell:2008hh, Fischer:2008kx, Fischer:2004tk, Parisi:2011wk, Fowler:2016bk, Fiel:2013ja, Owens2019bi, Owens2023lb, Lichter:2015gz}.
However, the indexes represent different concepts: the Information
Theory Index compares the diversity of local areas to the overall
diversity of a region
\citep{White:1986vx, Reardon:2002vt, Reardon:2004vl}, and the
Dissimilarity Index measures the difference between the local and
overall proportions of each group. Although studies have acknowledged
this difference, they commonly interpret values of the Information
Theory Index as carrying the same meaning as the Dissimilarity Index.
Given its characteristics, the Information Theory Index should be
interpreted as a measure of diversity and it should be used alongside
measures of segregation rather than as a substitute.

The Divergence Index is a decomposable measure of segregation and
inequality. As an index of racial residential segregation, it measures
how surprising the racial composition of local areas is given the
overall racial composition of the region. \label{R1_1a_p1} In this way,
it is similar to more conventional indexes, like the Dissimilarity
Index, that measure the ``evenness'' dimension of segregation
\citep{Massey:1988tq}. However, the Divergence Index also differs from
and thus complements such indexes by being additively decomposable. This
property is useful because it allows us to compare the segregation
occurring within and between groups or spatial units, such as nested
levels of geography. Additionally, the Divergence Index can be
calculated for both discrete and continuous distributions as well as for
joint distributions, such as income by race.

Below I introduce the Divergence Index in more detail. I begin by
comparing the concepts of segregation, inequality, and diversity, and
noting the key distinctions between them. I then illustrate what can be
learned through decomposition analyses of segregation with an empirical
example. Next, I provide a brief review of entropy measures of
diversity. I then introduce the Divergence Index and describe its unique
features. In the next section, I compare the mechanics of the Divergence
Index and Information Theory Index using stylized and empirical
examples.

Finally, I demonstrate the importance of the preserving the conceptual
distinction between segregation and diversity with a decomposition
analysis of the 100 largest metropolitan areas in the United States from
1990 to 2010. The analysis shows that the Information Theory Index has
tended to decrease overtime, particularly within cities, while the
Divergence Index has tended to increase, particularly within suburbs. I
explain how these differences can arise due to the the mechanics of the
indexes as well as different underlying population processes. The
Information Theory Index reveals the tendency for blocks to increase in
diversity at roughly the same pace as the region, but the Divergence
Index reveals the tendency for diversity promoting moves to reproduce or
intensify segregation. Despite a general tendency for segregation and
diversity to have a positive association --- the direction and strength
of the relationship varies across metros. In some metros increases in
diversity may reflect the transience of racial turnover that eventually
reproduces segregation, while in others it may be associated with the
gradual formation of stably integrated communities.

Analyzing segregation alongside diversity can enable richer, deeper
understandings of the rapidly but unevenly changing demographics of
societies such as the United States. The Divergence Index complements
existing measures by providing a distinct lens for understanding the
dimensions and dynamics of segregation.

\hypertarget{inequality-segregation-and-diversity}{%
\section{Inequality, Segregation, and
Diversity}\label{inequality-segregation-and-diversity}}

Social inequality and segregation are tightly coupled concepts.
Inequality refers to the uneven distribution of resources,
opportunities, or outcomes across a population (e.g., among individuals
or groups). Segregation refers to the uneven distribution of the
population across separate or distinct places, occupations, or
institutions. Hence inequality and segregation both involve the uneven
distribution of some quantity across units.

All measures of inequality and segregation have an implied or explicit
comparative reference that defines equality or evenness
\citep{Coulter:1989us}, such as the uniform distribution of income
across individuals or the random distribution of individuals across
neighborhoods. Measures evaluate the degree of inequality or segregation
for a given distribution by measuring it against the comparative
reference. For example, when a small portion of the population holds a
large share of income, the income distribution is unequal. When the
racial composition of neighborhoods differs widely across a city, racial
segregation is high. Segregation and inequality are relative and
relational concepts linked to social processes and mechanisms that
scholars often seek to understand.

The concept of diversity is another way to describe the composition of a
population, but it is not a relational concept. Diversity describes the
variety of ``types'' or groups in the population
\citep{Blau:1977ua, Page:2007tl, Page:2011wo}. Diversity indexes measure
the number of groups and in what proportion they are represented. Like
segregation and inequality, diversity represents the outcome of social
processes. Diversity has important relevance for sociological theories,
such as Blau's \citeyearpar{Blau:1977ua} theory of the structure of
social association, but it is less relevant than segregation and
inequality to examinations of hierarchical relations.

Unlike the relational concepts of segregation and inequality, diversity
is its indifferent to the specific groups that are over- or
under-represented in a population. For example, a neighborhood that is
75 percent White and 25 percent Black has the same amount diversity as a
neighborhood that is 25 percent White and 75 percent Black. Diversity
measures are only concerned with the variety or relative quantity of
groups, whereas inequality and segregation measures are concerned with
which groups (or which parts of a distribution) are over- and
under-represented. The two neighborhoods could have different
segregation values, and the level of segregation for each would depend
upon the city's composition.

Hierarchical relations are embedded within the concepts of segregation
and inequality. Groups may differ in their relative symbolic, economic,
or political advantage and measures of segregation and inequality can
preserve the distinction between groups. For example, measures of
segregation can distinguish between a neighborhood that is 75 percent
White and one that is 75 percent Black in a city that is 75 percent
White and 25 percent Black, and their group compositions imply different
levels of segregation within the city context.

In contrast, diversity indexes ``flatten fundamentally hierarchical
relations between groups. \ldots{} As an analytic concept, `diversity'
(i.e.~`heterogeneity') not only sidesteps issues of material and
symbolic inequalities, it masks the distinction between in-group and
out-group contact'' \citep[ p.~755]{Abascal:2015fq}. This becomes
problematic when a diversity index is interpreted as a measure of
segregation. Despite having the same amount of diversity, there may be
different social processes unfolding in neighborhoods with different
racial compositions and different levels of segregation. Segregation and
diversity are distinct concepts that allow us to answer different
questions about social processes and their outcomes \citep[see
also][]{Abascal2021pu}.

\hypertarget{decomposition-analysis-of-segregation}{%
\section{Decomposition Analysis of
Segregation}\label{decomposition-analysis-of-segregation}}

Scholars who use diversity as a measure of segregation are drawn to it,
I posit, in part because it permits decomposition analysis. In this
section, I provide an example of what can be learned through a
decomposition analysis that more traditional analyses do not provide,
and why decomposition is important for understanding segregation
patterns and processes. Decomposition, I seek to illustrate, is an
essential tool for understanding the multi-level dynamics of social and
spatial phenomena.

From 2000 to 2006, Westchester County received \$50 million in federal
housing funds, which required that the county affirmatively further fair
housing \citep{ADC2023wc}. In 2009, a federal judge ruled that the
county ``had misrepresented its efforts to desegregate overwhelmingly
white communities when it applied for the federal housing funds''
\citep{Santos2009jf}. Almost two-thirds of the county's Black and Latinx
residents lived in the county's four cities. Among the county's 39
villages and towns, only 8 had substantial Black or Hispanic
populations, while the other 31 remained White and segregated.

Westchester County settled the case and entered into a consent decree
(i.e.~a binding federal court order), which prohibited it from
``ignoring either the residential racial segregation that continues to
plague it, or the municipal resistance to affordable housing development
that stymies the possibility of changing those patterns''
\citep{ADC2023wc}. The federal monitor who later evaluated the county's
compliance with the consent decree pointed to low segregation within the
31 villages and towns as evidence of compliance. However, the levels of
segregation within these municipalities obscures the county-wide context
-- it ignores the segregation between municipalities and in the county
at large. A decomposition analysis of segregation, measured with the
Divergence Index, allows us to examine the segregation occurring both
within and between municipalities and how each contributes to the
county's overall segregation.

The additive decomposability property of the Divergence Index means that
the sum of segregation within areas and between areas is equal to the
overall segregation of the region. This allows us to answer questions
about the spatial organization of segregation by comparing the levels of
segregation within and between the areas. A drawback of using an index
that is not additively decomposable, such as the Dissimilarity Index,
for such a comparison is that it is possible to observe changes within
or between areas that leave the overall index unchanged.

Using the Divergence Index, the total segregation among the census
blocks\footnote{I use publicly available population data for census
  blocks --- the smallest unit of census geography --- from the 2010
  decennial census \citep{CensusSummary:bZv73ozJ}.} in Westchester
County is equal to the sum of two quantities: the segregation between
areas in the county and the average within-area segregation.
Between-area segregation represents how surprising the racial
composition of each area is given the county's overall racial
composition, and the within-area segregation represents how surprising
the racial composition of blocks within each area is given the area's
overall racial composition. I analyze segregation for municipalities
categorized into three groups (i.e., ``areas'') defined by the consent
decree: the 4 cities, the 8 municipalities excluded from the consent
decree (i.e., ``excluded areas''), and the 31 municipalities included in
the decree (i.e., ``included areas'') \citep{Beveridge2016rz}.

The decomposition results indicate that about a third of the segregation
in the county occurs between the cities, the excluded areas, and
included areas. The remaining two-thirds of the county's segregation
occurs among the blocks within each area, primarily within the cities
and excluded areas. This indicates that the blocks within the cities and
excluded areas are not very representative of each area's overall racial
composition. In contrast, as the monitor's report indicated, the
included areas have relatively low levels of within-area segregation ---
the meager Black and Latinx populations are distributed more evenly
across blocks. However, the included areas are responsible for most of
the segregation occurring between the areas in the county --- they have
a much higher proportion of White residents, compared to the other areas
or the county overall.

The segregation in Westchester County is occurring at two different
levels of spatial organization: among the areas within the county
(primarily in the included area), and among the blocks within each area
(primarily in the cities and excluded areas). Decisions about the
placement of affordable housing have important implications for reducing
segregation within the county, especially since Black and Latinx
residents occupy the majority of the county's subsidized and public
housing units \citep{Santos2009jf}. Increasing the number of affordable
housing units in the cities and excluded areas could \emph{increase}
between-area segregation by further concentrating Black and Latinx
residents in these areas. Placing affordable housing in the included
municipalities could \emph{decrease} between-area segregation by
creating affordable housing options across the county and affirmatively
furthering fair housing.

In the next section of the paper, I describe how entropy has been used
to measure diversity. I then introduce the Divergence Index as a
decomposable measure of segregation. Although the Divergence Index is
also an entropy-based measure, its formulation makes it well-suited to
measuring segregation and allows us to answer different types of
questions than a traditional entropy index or the Information Theory
Index.

\hypertarget{entropy-and-the-measurement-of-diversity}{%
\section{Entropy and the Measurement of
Diversity}\label{entropy-and-the-measurement-of-diversity}}

\hypertarget{entropy}{%
\subsection{Entropy}\label{entropy}}

Entropy is a commonly used measure in physics and information theory to
represent the randomness of a system or the information content of a
message
\citep{Coulter:1989us, Cover:2006ub, Shannon:1948iy, Theil:1967vj}.
\citet{Theil:1971jx} introduced the concept of entropy to the social
sciences as a measure of population diversity \citep[see
also][]{Reardon:2002vt, White:1986vx}. Entropy-based measures are
decomposable and have long been a staple of decomposition studies.

Entropy is the amount of information needed to describe a probability
distribution. If two outcomes are equally likely (e.g., a coin toss),
there is high uncertainly about what the outcome will be and high
entropy. If one outcome has a higher probability, there is less
uncertainty about what the outcome will be and lower entropy.\footnote{Entropy
  can be thought of as the uncertainty associated with the value of a
  random draw from a probability distribution. If an outcome has a
  probability of 100 percent, the entropy of the distribution is 0 ---
  there is no uncertainty. If there are two equally likely outcomes,
  such as with a fair coin toss, the entropy of each outcome (\(E_m\))
  is 1 and the average uncertainty (\(E\)) is 1, its maximum value. In
  other words, when two outcomes are equally likely, we have maximum
  uncertainty about what the outcome will be.} Entropy measures the
probability of an outcome (\(m\)) occurring, weighted by its probability
of occurrence (\(\pi_m\)). The entropy of each outcome (\(m\)) is
\(E_m=\log{\cfrac{1}{\pi_m}}\). Weighting each outcome by the
probability of its occurrence, the overall entropy is:
\[E=\sum_{m=1}^{M}{\pi_m\log{\cfrac{1}{\pi_m}}}.\]

The entropy equations can be defined using logarithms to any base. The
base of the logarithm defines the units of the index
\citep{Shannon:1948iy, Theil:1972vb}. Log base 2 (\(\log_2\))is
typically used in information theory, which gives results in units of
binary bits of information. It is common for inequality measures to use
the natural logarithm (\(\ln\)), which has the mathematical constant
(\(e\)) as its base. Following standard usage, we can define
\(0\log{0}=0\), because
\(\lim_{x\rightarrow0}{\left(x\log{x}\right)}=0\).

Interpreted as a measure of diversity, \(m\) indexes the groups (e.g.,
race or income group) in a population and the ``probability of an
outcome'' is the proportion of each group. If all individuals in a
population are associated with the same group, there is no diversity in
the population. There is no uncertainty about a randomly selected
individual's group membership, and entropy is equal to 0. On the other
hand, if individuals are evenly distributed among two or more mutually
exclusive groups, there is maximum diversity in the population, and
entropy is equal to 1.

The properties of entropy have been well documented
\citep[e.g.,][]{Cover:2006ub, Shannon:1948iy, Theil:1967vj}, and are
summarized in Table \ref{tab:evalTable}. It can be calculated for any
number of groups, and it has known upper and lower bounds with
substantive interpretations. Importantly, entropy and entropy-based
measures are additively decomposable (see Appendix C for equations). The
additivity of entropy comes from one of the properties of logarithms:
\(log{\left(\pi_1\cdot\pi_2\right)}=log{\left(\pi_1\right)}+log{\left(\pi_2\right)}\).

\hypertarget{the-information-theory-index}{%
\subsection{The Information Theory
Index}\label{the-information-theory-index}}

The Information Theory Index is an entropy-based measure developed by
\citet{Theil:1971jx} to study racial segregation in Chicago public
schools. The index has also been proposed as a measure of residential
segregation \citep{White:1986vx, Reardon:2004vl, Reardon:2002vt}. For a
single area (\(i\)), the Information Theory Index measures the extent to
which the area's entropy (\(E_i\)) is reduced below the region's entropy
(\(E\)), standardized by dividing by the region's entropy
\citep{Theil:1971jx}: \[H_i=\cfrac{E-E_i}{E}.\] Or, equivalently, it is
one minus the ratio of local diversity to overall diversity
\citep{Reardon:2002vt}: \[H_i=1-\cfrac{E_i}{E}.\] The region's index
value is the weighted average of \(H_i\) across all local areas:
\[H=1-\sum_{i=1}^{N}{\cfrac{\tau_iE_i}{TE}} ~ = ~ 1-\cfrac{\bar{E_i}}{E} \text{\hskip2em or \hskip2em} H=\sum_{i=1}^{N}{\cfrac{\tau_i}{T}H_i},\]
where \(T\) is the overall population count, and \(\tau_i\) is the
population count for area \(i\). \(H\) represents the relative reduction
in the average entropy of components (\(\bar{E_i}\)) below the maximum
attainable entropy (\(E\)) \citep{Theil:1971jx}. Or, equivalently, it is
one minus the ratio of average local diversity to overall diversity
\citep{Reardon:2002vt}.

The Information Theory Index is additively decomposable, meaning that we
can aggregate residential locations into districts and calculate the
index values within and between the districts in a region
\citep{Reardon:2002vt}. The sum of the within-district (\(H_{j}\)) and
between-district (\(H_{0}\)) components is equal to overall index for
the region: \begin{equation}\label{eq:H_decomp}  
H=H_0+\sum_{j=1}^{J}{\frac{T_jE_j}{TE}H_{j}}  
\end{equation}\\
where \(T\) is the overall population count and \(T_j\) is the
population count for district \(j\). The sum of the within-district
values is weighted by the relative population of the district
(\(\frac{T_j}{T}\)) and the relative diversity of the district
(\(\frac{E_j}{E}\)). Districts that are more populous and more diverse
contribute more to the overall index than less populous and less diverse
districts.

The Information Theory Index typically ranges between 0 and 1. A value
of 1 indicates that there is no diversity in local areas. A value of 0
indicates that all local areas are as diverse as the region. The minimum
value can be less than 0, and \citet{Reardon:2004vl} interpret negative
values of the index as indicating ``hyper-integration,'' which occurs
when localities are more diverse, on average, than the region as a
whole. In other words, groups are more equally represented in local
areas than in the overall population. Additional properties of the
Information Theory Index are summarized in Table \ref{tab:evalTable}.

In recent years, scholars have turned to the Information Theory Index as
the best alternative to the Dissimilarity Index in decomposition
analyses of segregation
\citep{Bischoff:2008bw, Farrell:2008hh, Fischer:2008kx, Fischer:2004tk, Parisi:2011wk, Fowler:2016bk, Fiel:2013ja, Owens2019bi, Owens2023lb, Lichter:2015gz}.
However, the indexes represent different concepts: the Information
Theory Index compares the diversity of local areas to the overall
diversity of a region
\citep{White:1986vx, Reardon:2002vt, Reardon:2004vl}, whereas the
Dissimilarity Index measures the difference between the local and
overall proportions of each group. Although studies have acknowledged
this difference, they commonly interpret the Information Theory Index as
carrying the same meaning as the Dissimilarity Index. Given its
characteristics, the Information Theory Index should be interpreted as a
measure of diversity and it should be used alongside measures of
segregation rather than as a substitute.

\hypertarget{the-divergence-index}{%
\section{The Divergence Index}\label{the-divergence-index}}

The Divergence Index addresses the need for a decomposable measure of
segregation. Decomposition is useful because it allows us to compare the
segregation occurring within and between groups or spatial units, such
as nested levels of geography. The additive decomposability property of
the index means that the sum of segregation within and between areas is
equal to the overall segregation of the region. This allows us to answer
questions about the spatial organization of segregation by comparing the
within and between components of the decomposition. In contrast, when
using indexes that are not additively decomposable, such as the
Dissimilarity Index, it is possible to observe changes within or between
areas while the overall index is unchanged. Decomposition is an
essential tool for understanding the multi-level dynamics of
segregation, as illustrated earlier in the case of Westchester County.

The Divergence Index is based on relative entropy, an information
theoretic measure of the difference between two probability
distributions \citep{Cover:2006ub}, also known as Kullback--Leibler (KL)
divergence \citep{Kullback:1987wm}. Relative entropy, shares many
properties with entropy, including being decomposable, but instead of
characterizing a single distribution, it compares one distribution to
another. The index can be used to measure inequality as well as
segregation.

The Divergence Index measures the difference between a distribution,
\(P\), and another empirical, theoretical, or normative distribution,
\(Q\).\footnote{The index measures the entropy of \(P\) relative to
  \(Q\), or the relative entropy of \(P\) with respect to \(Q\).} The
index represents the divergence of a model (\(Q\)) from reality (\(P\)).
It can be interpreted as a measure of surprise: How surprising are the
observations (\(P\)), given the expected value (\(Q\))? Or, how
surprising is an empirical distribution (\(P\)), given a theoretical
distribution (\(Q\))?

For discrete probability distributions \(P\) and \(Q\), the divergence
of \(Q\) from \(P\) is defined as:
\[D\left(P\parallel Q\right)=\sum_{m=1}^{M}{P_m\log{\cfrac{P_m}{Q_m}}},\]
where \(m\) represents groups or categories.\footnote{Using the
  properties of logarithms, the Divergence Index can also be written as
  \(D\left(P\parallel Q\right)=\sum_{m=1}^{M}{P_m\left(\log{P_m}-\log{Q_m}\right)}\).}
(Following standard usage, we can define \(0\log{0}=0\), because
\(\lim_{x\rightarrow0}{\left(x\log{x}\right)}=0\).) The \(Q\)
distribution defines the standard against which segregation or
inequality is measured. It should represent the expected state of
equality or evenness in the \(P\) distribution. \(Q\) can be
theoretically determined or empirically derived. For example, it can be
a standard probability distribution (e.g., a normal or uniform
distribution), a prior state of the \(P\) distribution, or the mean of
the observed data (\(P\)).

The Divergence Index has known upper and lower bounds with substantive
interpretations. The minimum value is 0, indicating no difference
between \(P\) and \(Q\). The maximum value can be less than or greater
than 1. The index can be standardized to have a range of 0 to 1 by
dividing by its maximum value for a given population. However,
standardizing the index transforms it from an absolute to a relative
measure of inequality and segregation, and negates several of its
desirable properties, including aggregation equivalence and
independence. (See \ref{tab:evalTable}.)

The Divergence Index is a non-symmetric measure of the difference
between two distributions \citep{Bavaud:2009fn}.\footnote{In contrast,
  entropy (\(E\)) is symmetric in \(P\left(x\right)\) and
  \(1-P\left(x\right)\).} The divergence of \(Q\) from \(P\) does not
necessarily equal the divergence of \(P\) from \(Q\). The asymmetry is
an intentional feature of the measure. As \citet{Bavaud:2009fn} states,
``the asymmetry of the relative entropy does not constitute a defect,
but perfectly matches the asymmetry between data and models'' (p.~57).

\label{R1_sp3} In the context of segregation measurement, this property
means that the divergence of each local area's composition from a
region's composition --- the divergence \(P\) from \(Q\) --- is not
necessarily the equal to the divergence of the region's composition from
the composition of local areas --- the divergence of \(Q\) from \(P\).
In the first case, the region's composition is the reference
distribution that the local compositions are compared to in evaluating
segregation. The region's composition represents the expected state of
the local compositions if there is no segregation in the region. In the
second case, each local area's composition is a reference distribution
that the region's composition is compared to. This second formulation is
inconsistent with practice of measuring segregation by evaluating how an
aggregate population is distributed across units. It is possible to
calculate a symmetric version of the index as the sum of
\(D\left(P\parallel Q\right)\) and \(D\left(Q\parallel P\right)\), but
such an index would likewise not represent the common conceptualization
of segregation as the uneven distribution of the population across
separate or distinct places, occupations, or institutions --- the
concept of segregation that motivates this paper.

One of the unique features of the Divergence Index is that it can be
calculated for either discrete distributions (relative entropy) or
continuous distributions (differential relative entropy)
\citep{Cover:2006ub}.\\
For continuous probability distributions \(P\) and \(Q\), the divergence
of \(Q\) from \(P\) is defined as:
\[D\left(P\parallel Q\right)=\int{p\left(x\right)\log{\cfrac{p\left(x\right)}{q\left(x\right)}}dx},\]
where \(p\) and \(q\) represent the probability densities of \(P\) and
\(Q\). The desirable properties of both relative entropy and
differential relative entropy have been well documented
\citep[e.g.,][]{Bavaud:2009fn, Cover:2006ub}. (Table \ref{tab:evalTable}
summarizes the properties of the Divergence Index.)

\hypertarget{measuring-residential-segregation-with-the-divergence-index}{%
\subsection{Measuring Residential Segregation with the Divergence
Index}\label{measuring-residential-segregation-with-the-divergence-index}}

As a measure of residential segregation, the Divergence Index can be
calculated for discrete distributions, such as the proportion of each
race group, or continuous distributions, such as the distribution of
income. In this section, I focus on using the Divergence Index to
measure residential segregation for discrete groups.

To study residential segregation, the Divergence Index measures the
difference between the overall proportion of each group in the region
(e.g., a city or metropolitan area) and the proportion of each group in
local areas within the region (e.g., census tracts). The overall
proportion of each group in the region is the reference distribution
(\(Q\)), which represents the expected local proportion of each group if
there is no segregation. The index asks: how surprising is the
composition of local areas given the overall population of the region?
If there is no difference between the local proportions of each group
and the overall proportions, then there is no segregation in the region.
More divergence between the overall and local proportions indicates more
segregation.

\label{R1_1b_p3} Like the Dissimilarity Index, the Divergence Index
measures how evenly the population of each group is distributed across
locations in the region. However, the Dissimilarity Index uses a linear
function to evaluate segregation and the Divergence Index uses a
logarithmic function. This means that, with the Dissimilarity Index, any
departure from evenness is treated equally, whereas the Divergence Index
evaluates small departures from evenness as contributing proportionally
less to segregation than larger departures from evenness.\footnote{The
  greater the divergence of \(Q\) from \(P\), the lower the probability
  of observing the local proportions (\(P\)) if there is no segregation
  in the region (\(Q\)).} For example, in a city with two equally sized
groups, the local Divergence Index values for neighborhoods with a 5
percent difference in their group compositions will be more similar if
their compositions are closer to the city's 50-50 composition (e.g.,
45-55 and 40-60) than if their compositions are very different from the
city's composition (e.g., 5-95 and 0-100). In other words, the further
from evenness the local composition is, the more surprising each unit of
departure becomes. With the Dissimilarity Index, however, any departure
from evenness contributes equally to the segregation value.\footnote{The
  Dissimilarity Index measures the deviation of each location's
  population composition from the overall population composition. Or,
  equivalently, it measures how evenly the population of each group is
  distributed across a region. It is calculated as the absolute
  difference between the proportion of groups \(A\) and \(B\) in the
  \(i\)\textsuperscript{th} location, summed over all locations and
  divided by 2:
  \[DI=\cfrac{1}{2}\sum_{i=1}^{N}{\left|\cfrac{\tau_{iA}}{T_A}-\cfrac{\tau_{iB}}{T_B}\right|},\]
  where \(\tau_{iA}\) is group \(A\)'s population count in location
  \(i\) and \(T_A\) is the total population of group \(A\), and likewise
  for group \(B\). If group A and B are distributed across locations in
  the same proportions, then there is no segregation. Segregation is
  measured as the extent to which the spatial distribution of group B
  deviates from that of group A. Although the index is typically used to
  measure segregation for two mutually exclusive groups, it can be
  rewritten to measure the segregation of multiple groups
  \citep{Morgan:1975vu, Sakoda:1981uw, Reardon:2002vt}. It is also used
  to measure inequality, known as mean relative deviation
  \citep{Reardon:2002vt}. The Dissimilarity Index has a number of
  notable limitations, which have been well documented and are
  summarized in Table \ref{tab:evalTable}.}

The Divergence Index for location \(i\) is:
\[D_i=\sum_{m=1}^{M}{\pi_{im}\log{\cfrac{\pi_{im}}{\pi_m}}},\] where
\(\pi_{im}\) is group \(m\)'s proportion of the population in location
\(i\), and \(\pi_m\) is group \(m\)'s proportion of the overall
population. If a location has the same composition as the overall
population, then \(D_i=0\), indicating no segregation.

\label{rev_prox1} To measure segregation spatially, a proximity-weighted
contribution from nearby areas is included in each location's
population. This creates overlapping local environments or
\emph{ego-centric neighborhoods}
\citep{Roberto:2018jw, Lee:2008gm, Reardon:2008wa, Reardon:2009kq}. In
the equation for the Divergence Index, we would replace \(\pi_{im}\)
with \(\tilde{\pi}_{rim}\), which is group \(m\)'s proportion of the
spatially weighted population within a given distance \(r\) of location
\(i\) \citep[for examples, see][]{Roberto:2018jw}:
\begin{equation}\label{eq:Di_spatial}  
\tilde{D}_{ri} = \sum_{m=1}^{M}{\tilde{\pi}_{rim}\log{\cfrac{\tilde{\pi}_{rim}}{\pi_m}}}.  
\end{equation}\\
A proximity function, such as a uniform or distance-decay function,
weights the relative contribution of distant vs.~nearby locations to the
spatially weighted population of each location. The proximity weighted
population composition for location \(i\) is calculated by dividing the
proximity weighted population count of each group in \(i\)'s local
environment, \(\tilde{\tau}_{rim}\), by the total proximity weighted
population count in the local environment, \(\tilde{\tau}_{ri}\):\\
\begin{equation}\label{eq:Pi_rim}  
  \tilde{\pi}_{rim} = \frac{\tilde{\tau}_{rim}}{\tilde{\tau}_{ri}}.  
\end{equation}\\
The value of \(\tilde{\pi}_{rim}\) represents the population composition
experienced by the residents of location \(i\) in their local
environment with a reach of \(r\), where the populations of other
locations are weighted according to the proximity function
\citep{Roberto:2018jw}.

Overall segregation in the region is the population-weighted average of
the divergence for all locations:
\[D=\sum _{i=1}^{N}{\frac{\tau_i}{T}D_i},\] where \(T\) is the overall
population count, and \(\tau_i\) is the population count for location
\(i\). If all locations have the same composition as the overall
population, then \(D=0\), indicating no segregation in the region.

The Divergence Index is additively decomposable, meaning that we can
aggregate residential locations into districts and calculate the
segregation occurring within and between the districts in a region. The
sum of the within and between components of segregation is equal to
overall segregation for the region. For example, to measure residential
segregation for districts within a city, we rewrite the Divergence Index
as the sum of between-district segregation and the average
within-district segregation. The average within-district segregation for
district \(j\) is:
\[D_{j}=\sum_{i\in S_j}{\frac{\tau_i}{T_j}\sum_{m=1}^{M}{\pi_{im}\log{\cfrac{\pi_{im}}{\pi_{jm}}}}},\]
where \(S_j\) is the set of locations in district \(j\). The reference
distribution, \(\pi_{jm}\), is the population composition of district
\(j\), which is calculated as the population-weighted average of the
group proportions for all localities (\(i\)) within the district:
\(\pi_{jm}=\sum_{i\in S_j}{\cfrac{\tau_i}{T_j}\pi_{im}}\), where \(T_j\)
is the population count for district \(j\). The between-district
segregation is:
\[D_{0}=\sum_{j=1}^{J}{\cfrac{T_j}{T}\sum_{m=1}^{M}{\pi_{jm}\log{\cfrac{\pi_{jm}}{\pi_{m}}}}}\]
Total segregation is the sum of the between-district segregation
(\(D_{0}\)) and the average within-district segregation (\(D_{j}\)):
\begin{equation}\label{eq:D_decomp}  
D=D_0+\sum_{j=1}^{J}{\frac{T_j}{T}D_{j}}.  
\end{equation}\\
The within component of segregation is a population-weighted sum of the
within-district segregation for all districts. In contrast, the
decomposition of the Information Theory Index, \(H\), applies two
weights in calculating the within component: it is a population
\emph{and} relative diversity weighted sum of the within-district index
values for all districts (see Equation \ref{eq:H_decomp}).

\hypertarget{related-indexes}{%
\subsection{Related Indexes}\label{related-indexes}}

Several other measures of inequality and segregation have been derived
from relative entropy. Theil's index of income inequality is a special
case of relative entropy, which measures the difference between the
observed distribution of income and the mean \citep{Theil:1967vj}. The
theoretical state of equality is one in which everyone's income is equal
to the mean. Theil's index belongs to the generalized entropy class of
measures, which also includes mean log deviation, half the coefficient
of variation, and the Atkinson Index
\citep{Cowell:1981vu, Cowell:1980un, Shorrocks:1980hi, Shorrocks:1984ip, Cowell:1980tu, Breen:2011ha}.\footnote{The
  Theil Index is approximately equivalent to Atkinson's inequality index
  with weights that are close to 0 in its social welfare function
  \citep{Schwartz:1980tb}.} The Divergence Index can likewise be used to
compare a distribution to a single value (see Appendix D), but also
provides the flexibility to holistically compare two distributions.

The use of relative entropy was incorporated into the ``relative
distribution'' method for measuring inequality \citep{Handcock:1999wh}.
The relative distribution method compares distributions rather than
summarizing their individual shapes, as with the Theil Index. The method
also includes the median relative polarization index, which summarizes
changes in the relative distribution. Relative distribution measures are
used to analyze specific distributional shifts in income and have been
reviewed in detail elsewhere
\citep{Liao:2002wk, Handcock:1999wh, Handcock:1998gj, Hao:2010vr}.

Additional applications of relative entropy in sociology have included a
summary measure of racial disparity that compares the distribution of
income for White and Black households \citep{Bloome:2014ga} and a
measure of educational disparities in adult mortality
\citep{Sasson:2016dl}. In economics, divergence is used to study
industrial localization and agglomeration \citep[e.g.,][]{Mori:2005wu}.
More generally, divergence underlies popular statistical methods of
model selection, including the Akaike Information Criterion (AIC)
\citep{Akaike:1974ih}.

A related information theoretic measure called mutual information is a
special case of relative entropy \citep{Cover:2006ub}. Whereas relative
entropy measures the difference between two probability distributions,
mutual information measures the amount of information shared between two
random variables. It quantifies the reduction in uncertainty about one
random variable, given knowledge about another random variable.

In the social sciences, a mutual information index has been proposed as
a measure of school segregation
\citep{Frankel:2011ex, Mora:2011ii, Mora2009ip}, residential segregation
\citep{Elbers2023ms, Kramer2018dn}, and occupational segregation
\citep{Alonso-Villar.2010}. In these cases, the random variables
represent groups (e.g., racial or income groups) and locations (e.g.,
schools or census tracts). For example, the index can measure how much
we know about a person's race by knowing where they go to school.
Because it is a symmetric measure, it also tells us how much we know
about where a person goes to school by knowing their race.

\hypertarget{differences-between-the-divergence-index-and-mutual-information.}{%
\subsubsection{Differences between the Divergence Index and Mutual
Information.}\label{differences-between-the-divergence-index-and-mutual-information.}}

Mutual information can be written as the relative entropy between a
joint distribution and the product of its marginal distributions, and in
many cases it can produce the same segregation value as the Divergence
Index. However, the indexes differ in important ways, including how they
are formulated and their potential use cases.

Mutual information is calculated using a joint probability distribution
and its marginal distributions, each of which sum to 1. Because of this,
the comparative reference of segregation cannot be independently
specified. In contrast, the Divergence Index measures the difference
between two probability distributions, and the distribution that
represents the comparative reference can be a marginal distribution or
another theoretically relevant distribution.

Segregation indexes typically use conditional probabilities to measure
segregation rather than joint probabilities
\citep[see][]{Grannis:2002wv}. For example, datasets used for measuring
residential segregation are structured so that each row represents a
neighborhood and the proportion of each group in a neighborhood's
population are included in the columns. Each row of the dataset --- each
neighborhood --- sums to 1. A joint probability distribution used for
the mutual information index would have the same basic structure, with
neighborhoods in rows and groups in columns, but proportions for the
entire dataset would sum to 1, so that each cell represents the
proportion of the region's population in a given neighborhood and group.

This distinction matters in cases when the relevant comparative
reference for segregation --- the \(Q\) distribution --- is a
counterfactual or theoretical distribution, or any time the aggregate
proportions used as the comparative reference are not equal to the
product of the marginals of a joint distribution. This can occur when
using non-exclusive subunits, such as studying overlapping relationships
in personal and business networks \citep[e.g.,][]{Smith2016tt}, or when
measuring spatial segregation or local segregation for a subset of
locations. In such cases, we may wish to use the composition of the
network or region as the \(Q\) distribution rather than the product of
the marginals of a joint distribution.

For example, when measuring spatial segregation with overlapping local
environments around each location in a city
\citep[e.g.,][]{Roberto:2018jw}, the populations of the local
environments are not mutually exclusive --- they include the residents
of the location itself, as well as the populations of other nearby
locations within a particular distance. Such an approach allows us to
compare the level of segregation for differently sized local
environments and examine the geographic scale of segregation patterns.
We would use the region's composition as the \(Q\) distribution in order
to evaluate the difference between the local and regional proportions as
the size of local environments changes. As the local environments
increase in size and include more and more of the region's population,
they would eventually begin to converge on the regional proportions.

The overlapping nature of the local environments means that if we used
the product of the marginals of the joint distribution as the \(Q\)
distribution, a single location's population could be counted multiple
times. Thus the aggregate population of all local environments may not
be equal to the region's population composition, and the population of
more central locations may be counted more heavily than outlying
locations in the aggregate population if they fall within more
locations' local environments. In such cases, it is preferable to use
the Divergence Index and specify the probability distributions being
compared, rather than using mutual information, which necessitates using
a joint probability distribution and its marginal distributions.

\hypertarget{comparing-the-divergence-index-and-information-theory-index}{%
\section{Comparing the Divergence Index and Information Theory
Index}\label{comparing-the-divergence-index-and-information-theory-index}}

In this section, I compare the mechanics of the Divergence Index and
Information Theory Index using stylized and empirical examples. The
Divergence Index and Information Theory Index share many desirable
properties, particularly their decomposability. A key difference between
them is the concepts they measure. The Divergence Index measures
segregation and inequality, and the Information Theory Index measures
diversity. The concept of diversity concerns the variety or relative
quantity of groups in a population, whereas segregation concerns the
degree to which specific groups are over- or under-represented in the
local areas compared to the region's composition.\footnote{Measures of
  diversity can not distinguish between a setting in which the
  proportion of a minority group and a majority group match their
  proportions in the overall population, and one in which the
  proportions of the minority and majority groups are swapped.} I
provide this comparison between the indexes because the Information
Theory Index has been widely used in decomposition studies of
segregation.

\hypertarget{equivalence-between-the-overall-indexes}{%
\subsection{Equivalence between the Overall
Indexes}\label{equivalence-between-the-overall-indexes}}

The Information Theory Index, \(H\), measures the ratio of local
diversity to overall diversity. Whereas the Divergence Index, \(D\),
measures the difference between the local and overall group proportions.
It is possible to derive an equivalence between \(H\) and \(D\) at the
aggregate level of a city or region, but only if overall entropy (\(E\))
is nonnegative and greater than or equal to the average local entropy
(\(\bar{E_i}\)): \(0 \le E \ge \bar{E_i}\). However, no such equivalence
exists at the local-level for locations or districts within a city or
region.

If both conditions hold then, we can derive the equivalence between
\(H\) and \(D\) by first rewriting the equation for \(D\) as:
\(D=E-\bar{E_i}\) (see Appendix A). Recall that we can write the
equation for \(H\) as: \(H=\cfrac{E-\bar{E_i}}{E}\). From this, we can
derive the equivalence as:
\[H=\cfrac{D}{E} \text{\hskip1em and \hskip1em} D=HE.\] \(H\) is
equivalent to \(D\) standardized by \(E\), or the ratio of \(D\) to
\(E\). Next, I describe the conditions that lead \(E\) to be negative or
less than the average local entropy --- if either occurs, then the
equivalence does not apply.

\hypertarget{overall-entropy-is-negative.}{%
\subsubsection{Overall Entropy is
Negative.}\label{overall-entropy-is-negative.}}

The entropy of a discrete distribution is always nonnegative, however
Cover and Thomas \citeyearpar[244]{Cover:2006ub} show that the entropy
of a continuous distribution (called ``differential entropy'') can be
negative. For example, the differential entropy of a uniform
distribution \(U\left( 0,a \right)\) is negative for \(0<a<1\). This
occurs because the density of the distribution is \(\cfrac{1}{a}\) from
\(0\) to \(a\), and
\[E=-\int_{0}^{a}{\cfrac{1}{a}\log{\cfrac{1}{a}}}~dx=\log{a}.\] Because
\(a<1\), therefore \(\log{a}<0\). In contrast, both relative entropy and
differential relative entropy (the discrete and continuous versions of
the Divergence Index) are always nonnegative \citep{Cover:2006ub}.

\hypertarget{average-local-entropy-is-greater-than-overall-entropy.}{%
\subsubsection{Average Local Entropy is Greater than Overall
Entropy.}\label{average-local-entropy-is-greater-than-overall-entropy.}}

Average local entropy (\(\bar{E_i}\)) can be greater than overall
entropy (\(E\)) if three conditions hold: if the overall population is
not maximally diverse (i.e.~the groups have different population sizes),
if any subunits have more diversity than the overall population (e.g.,
if there are local areas where groups are more similar in size than in
the overall population), and if the subunits are not mutually exclusive.

\label{rev_prox2} The first two conditions are quite common when
measuring segregation. The third condition --- non-exclusive subunits
--- arises when measuring segregation spatially. Spatial segregation
measures, including the spatial version of the Divergence Index in
equation \ref{eq:Di_spatial} \citep{Roberto:2018jw}, include a
proximity-weighted contribution from nearby areas in each location's
population. This creates overlapping local environments or
\emph{ego-centric neighborhoods}
\citep{Lee:2008gm, Reardon:2008wa, Reardon:2009kq}, which are not
mutually exclusive: they include the residents of the location itself,
as well as the populations of other nearby locations within a particular
distance. Non-exclusive subunits are also common in social network
analysis, such as studying overlapping relationships in personal and
business networks \citep[e.g.,][]{Smith2016tt}.

When the three conditions listed above occur, then average local entropy
(\(\bar{E_i}\)) can be greater than overall entropy (\(E\)), and \(E\)
can not be used to derive the equivalence between the Information Theory
Index and the Divergence Index. Moreover, when \(\bar{E_i}\) is greater
than \(E\), then the Information Theory Index will be
negative.\footnote{It is possible to observe nonnegative values of \(H\)
  when \(E\) is negative, but only if \(\bar{E_i}\) is also negative.}

\hypertarget{comparing-the-local-indexes}{%
\subsection{Comparing the Local
Indexes}\label{comparing-the-local-indexes}}

\begin{figure}[h]
  \begin{subfigure}{0.329\linewidth}
    \includegraphics[width=\linewidth, center]{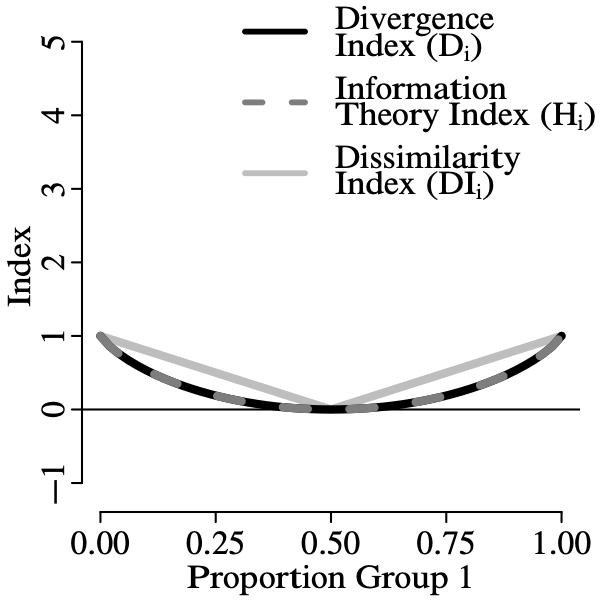}
    \caption{City A\\ Overall Group\\ Proportions: 0.5, 0.5}
    \label{fig:cityA}
  \end{subfigure}  
  \begin{subfigure}{0.329\linewidth}
    \includegraphics[width=\linewidth, center]{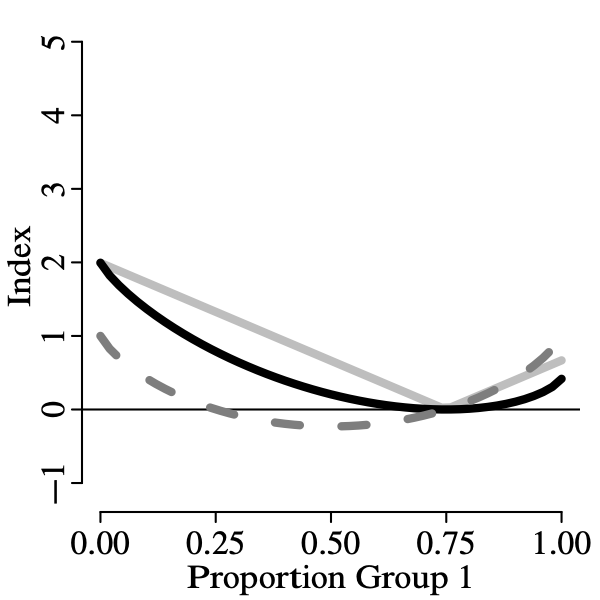}
    \caption{City B\\ Overall Group\\ Proportions: 0.75, 0.25}
    \label{fig:cityB}
  \end{subfigure}  
  \begin{subfigure}{0.329\linewidth}
    \includegraphics[width=\linewidth, center]{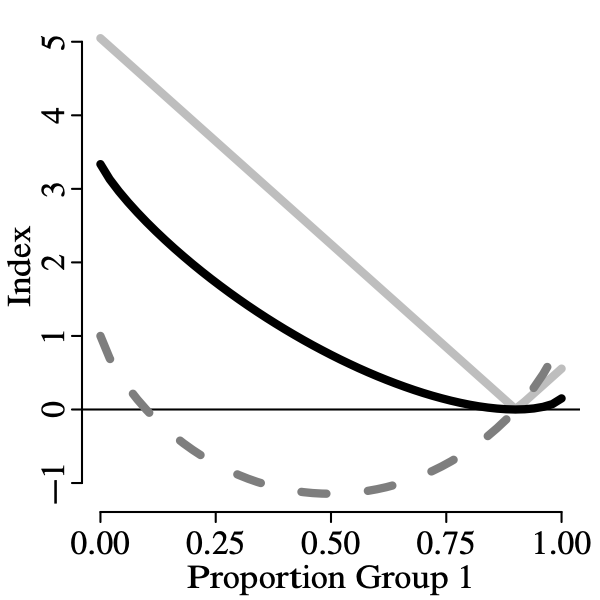}
    \caption{City C\\ Overall Group\\ Proportions: 0.9, 0.1}
    \label{fig:cityC}
  \end{subfigure}  
  \captionsetup{width=\linewidth}
  \caption{Comparing Local Values of the Divergence Index and\\ Information Theory Index in Three Hypothetical Cities}
  \label{fig:compareHD}
\end{figure}

To illustrate the similarities and differences between the Divergence
Index and Information Theory Index, Figure \ref{fig:compareHD} compares
the functional form of the local indexes for three hypothetical cities.
For the sake of the illustration, the two conditions listed above are
both satisfied --- overall entropy in the cities is positive, and
average local entropy is not greater than overall entropy --- and an
equivalence exists between the city-level indexes, though not the local
indexes. For reference, Figure \ref{fig:compareHD} also includes the
functional form of local values for the Dissimilarity Index.

Each city is divided into mutually exclusive local areas, and there are
two groups in the cities' populations. The proportion of each group
varies across cities: 50-50 in city A, 75-25 in city B, and 90-10 in
city C. The horizontal axes in Figure \ref{fig:compareHD} show the
proportion of group 1 in the local areas within each city. The vertical
axes show the index values for local areas within the city across the
full range of possible values for the local proportions of group 1. The
dark solid lines plot the local index values for \(D_i\), the dashed
lines plot the local index values for \(H_i\), and for reference, the
light solid lines plot the local index values for the Dissimilarity
Index \(DI_i\).

\vspace{1ex}
\centerline{[Figure \ref{fig:compareHD} about here.]}

The minimum and maximum values of \(D_i\) and \(H_i\) vary across the
three hypothetical cities in Figure \ref{fig:compareHD}. Local values of
the Divergence Index, \(D_i\), take their minimum value, which is always
0, when the local population composition is the same as the overall
composition of the city. \(D_i\) reaches its maximum value when a city's
minority group is 100 percent of the local population. In a city where
two groups are equally represented, like city A, it is just as
surprising to observe a location where 100 percent of the residents are
in group 1 as a location where 100 percent of the residents are in group
2. However, when there is a large majority group, as in cities B and C,
it is more surprising to observe a location where all residents are in
the minority group than a location where all residents are in the
majority group. Further, it is more surprising to observe a location
where all residents are in the minority group in a city C with a 10
percent minority population than in a city B with a 25 percent minority
population. This is demonstrated in Figure \ref{fig:compareHD} by
comparing the local value of the Divergence Index in cities A, B, and C
when the local proportion of the majority group (group 1) is 0.

By contrast, local values of the Information Theory Index, \(H_i\),
reach their maximum value when \emph{any group} is 100 percent of the
local population, regardless of the city's population composition.
\(H_i\) equals 0 when local diversity is the same as the city's
diversity, regardless of whether any group is over- or under-represented
in the local population. For example, Figure \ref{fig:cityB} shows that
\(H_i=0\) when the proportion of group 1 in the local population is
either 0.25 or 0.75, even though the proportion of group 1 in the city
is 0.75. The functional form of \(H_i\) for a two-group index is always
symmetric around 0.5 --- even proportions of each group --- regardless
of the overall population composition. \label{R1_1b_p1} In contrast, the
Divergence Index and Dissimilarity Index are symmetric only if there are
even proportions of each group in the overall population, otherwise
local values of both indexes are zero only if the local proportion of
each group is the same as the overall population.

\(H_i\) takes its minimum value, which is typically less than zero, when
a local area has an even mix of groups, regardless of the city's
diversity. The minimum value of \(H_i\) is a decreasing function of the
city's overall diversity. (Recall that \(H_i\) is 1 minus the ratio of
local diversity to overall diversity.) Given the same level of local
diversity, the value of \(H_i\) will be lower in a city with a less
overall diversity than in a city with more overall diversity. This is
demonstrated in Figure \ref{fig:compareHD} by comparing across cities.
The inflection point, or minimum value, of the function for \(H_i\) is 0
in city A where there is an even mix of groups, slightly negative in
city B, and even more negative in city C, which is the least diverse
city with a 90-10 mix of groups.

If local areas are marginally more diverse, on average, than the overall
population, then \(H\) will be negative.\footnote{Negative values of
  \(H\) occur when \(\bar{E_i}\) is greater than \(E\). (Recall that
  \(H=1-\cfrac{\bar{E_i}}{E}\).) In a previous section, I explained the
  conditions under which this occurs.} \citet{Reardon:2004vl} interpret
negative values of \(H\) as indicating ``hyper-integration'' --- each
group is more equally represented in local areas, on average, than in
the overall population. In contrast, \(D\) and \(D_i\) are never
negative \citep{Cover:2006ub}.

The values of the indexes are the \emph{same} when there is an even mix
of groups in the city population, as in city A (Figure \ref{fig:cityA}).
If the proportion of each group in local areas is the same as the city
proportions, then both indexes equal zero. If all local areas are
monoracial, such that each group is either 100 percent or 0 percent of
the local population, then both city-level indexes reach their maximum
value. If the proportion of each group varies across local areas, then
the measures would each find some degree of segregation or relative
homogeneity. Moreover, the values of both indexes will be the same in
the rare case that the overall population is maximally diverse.

The difference between the indexes is greatest when there is a small
minority group in the population. At the extreme, if there is only one
group present in the city and all local areas are monoracial, \(D\) and
\(H\) give \emph{opposite} results. \(H\) would show that the city is
maximally homogenous (all \(H_i=1\) and \(H=1\)) because there is no
diversity in either the local areas or the city.\footnote{Technically,
  \(H\) is undefined if there is only one group in the population,
  because \(H=1-\cfrac{0}{0}\). If there are two groups in the
  population, the limit of \(H\) as the minority group's population
  count approaches 0 (and \(E\) and \(\bar{E_i}\) approach 0) is 1.} In
contrast, \(D\) would find that the city is not at all segregated (all
\(D_i=0\) and \(D=0\)), because there is no difference between the
composition of local areas and the city as a whole --- each local area
is a microcosm of the city. \label{R1_sp6} Although there does not seem
to be a reason to compute the indexes when there is only one group in a
city, this situation can arise when, for example, computing the indexes
for multiple cities where a particular ethnoracial group may be present
in some cities but not others. The typical approach with \(H\) (or the
Dissimilarity Index) is to restrict the sample of cities to those with a
population of at least 1,000 for all groups. This selection criteria is
not necessary when using \(D\) --- the groups can be of any size or
absent in some contexts but not others.

\label{R1_sp7_p1} Table \ref{tab:nbhdsegdiv} lists local index values
for hypothetical neighborhoods in city B that each have the same
population count but different group proportions. The table shows the
correspondence between local and overall values of the indexes, and
illustrates the conditions when they indexes give disparate results.

The value of \(H_i\) is 0 in the neighborhood with 75-25 group
proportions, the same as city B, and 1 in the neighborhood with 100-0
group proportions. In contrast, the values of \(D_i\) are 0 and .415 in
these two neighborhoods. The difference between these neighborhoods is
greater for \(H_i\) than \(D_i\).As noted earlier, the value of \(H_i\)
is always 1 in neighborhoods with only majority group residents,
regardless of the region's group proportions. (See Figure
\ref{fig:compareHD}.) In contrast, the value of \(D_i\) adapts to
represent the degree of difference between the local and overall group
proportions.

An implication of this is that the overall value of \(H\) is always
greater than \(D\) --- when measured for two groups --- and the
difference is greater in regions where 1) the overall group proportions
are more unequal, and 2) the local group proportions tend to be more
extreme (i.e., all one group or the other).

\vspace{1ex}
\centerline{[Table \ref{tab:nbhdsegdiv} about here.]}

\label{R1_1b_p2} The most notable difference between the Dissimilarity
Index (\(DI_i\)) and both \(D_i\) and \(H_i\) is that local values of
\(DI_i\) follow a linear function, whereas both \(D_i\) and \(H_i\) are
logarithmic functions of the local proportions. Local values of the
Dissimilarity Index (\(DI_i\)) are always greater than or equal to
\(D_i\). Local values of the \(DI_i\) are greater than or equal to
\(H_i\), except in locations where the majority group approaches 100
percent of the local population.

\begin{figure}[h]
  \includegraphics[width=0.5\linewidth, center]{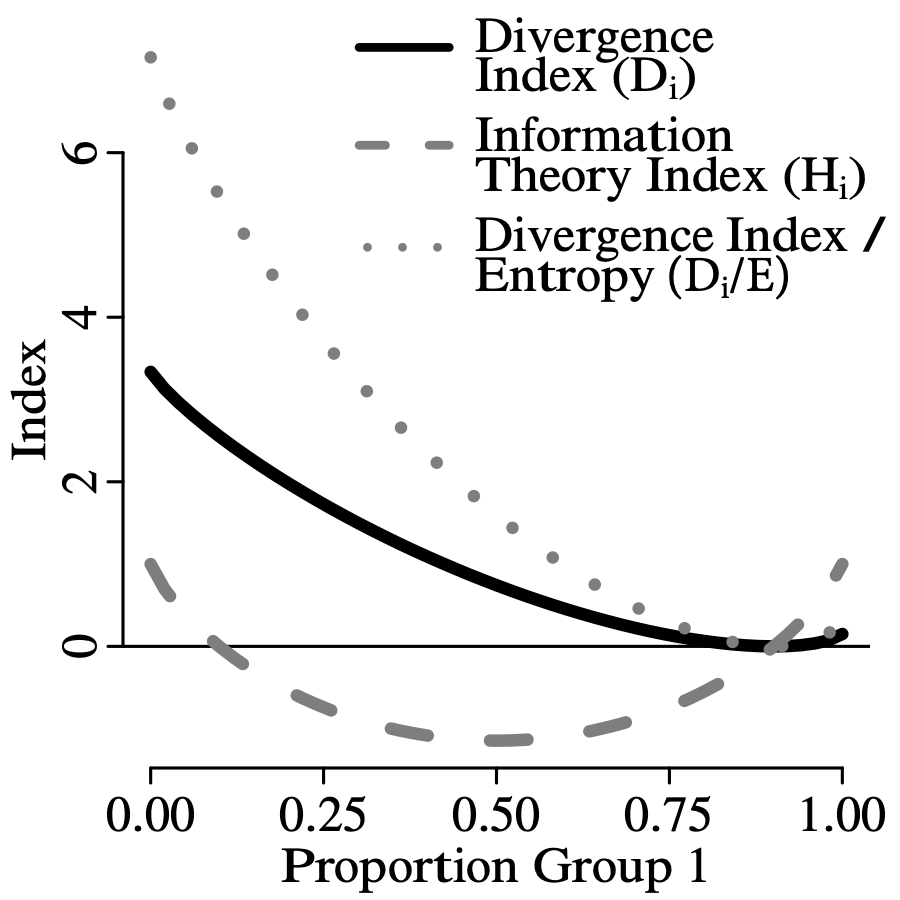}
  \captionsetup{width=\linewidth}
  \caption{Comparing Local Values of the Indexes in City C}
  \label{fig:compareHDE}
\end{figure}

In the previous section, I explained the conditions under which \(E\)
can be used to derive an equivalence between \(D\) and \(H\) and noted
that no such equivalence exists for the local values of the indexes. If
the local values of the Divergence Index, \(D_i\), are divided by
overall entropy, \(E\), the resulting values have a functional form that
is similar to \(D_i\) but with the same population-weighted mean as
\(H_i\). Figure \ref{fig:compareHDE} uses the population and index
values of city C to illustrate this point. The values for \(D_i\) and
\(H_i\) are unchanged from Figure \ref{fig:cityC}, and the dotted line
represents the values of \(D_i/E\).

\vspace{1ex}
\centerline{[Figure \ref{fig:compareHDE} about here.]}

Despite having the same population-weighted mean as \(H_i\), \(D_i/E\)
does not share the unique symmetry property of \(H_i\), where regardless
of the city's composition, \(H_i\) always takes its minimum value when
there is an even mix of groups in the local population (e.g., 50-50 for
a two-group index) and always takes its maximum value when \emph{any
group} is 100 percent of the local population. As with \(D_i\),
\(D_i/E\) takes its minimum value when local areas have the same
composition as the city and reaches its maximum value when the city's
minority group is 100 percent of the local population. As illustrated in
Figure \ref{fig:compareHDE}, the values of \(D_i/E\) are always greater
than \(D_i\) when \(E<1\), such as with a two-group index when there are
uneven group proportions. This is also true for the overall values of
\(H\) --- \(H>D\) when \(E<1\) --- but not for the local values.

\(D_i/E\) may be a better alternative to \(H_i\) for local values.
However, using \(D/E\) does not resolve an additional issue with the
decomposition of \(H\), where the within components for each subarea are
not equal to the population-weighted index values for each subarea
alone. \(D/E\), like \(H\), has an additional diversity weight that must
be applied. As a result, subareas that are more populous and more
diverse contribute more to the overall index than less populous and less
diverse districts. I will illustrate this point in greater detail in the
following section.

\hypertarget{decomposing-the-divergence-index-and-information-theory-index}{%
\subsection{Decomposing the Divergence Index and Information Theory
Index}\label{decomposing-the-divergence-index-and-information-theory-index}}

In this section, I decompose racial residential segregation and relative
homogeneity in the Detroit, MI, metropolitan area using the Divergence
Index and Information Theory Index. I provide this comparison because
previous studies have used the Information Theory Index as the best
alternative to the Dissimilarity Index for decomposing segregation
within and between communities, municipalities, or school districts
\citep{Bischoff:2008bw, Farrell:2008hh, Fischer:2008kx, Fischer:2004tk, Parisi:2011wk, Fowler:2016bk, Fiel:2013ja, Owens2019bi, Owens2023lb, Lichter:2015gz}
and commonly interpret the Information Theory Index as a measure of
segregation. The analysis demonstrates key differences in the
calculation and interpretation of decomposition results for the
Divergence Index and Information Theory Index, including the weighting
of the within-subarea values and the calculation of local values in the
between-subareas component. This section focuses on the mechanics of the
indexes, using Detroit as an example, whereas the following section has
an empirical focus --- describing changes over time in segregation and
diversity in the 100 largest U.S. metros.\footnote{All analysis were
  conducted using R software \citep{RCoreTeam2023le}.}

The Detroit metropolitan area is commonly cited as one of the most
racially segregated places in the U.S. A large majority of the city's
residents are Black, while the surrounding area's population is
predominantly White.\footnote{Using the U.S. Census Bureau's categories
  of race and ethnicity, I define two mutually exclusive ethnoracial
  groups for this analysis: non-Hispanic Black (``Black'') and
  non-Hispanic White (``White'').} To better understand the regional
dynamics of segregation, I measure Black-White segregation in the
Detroit metro area with the Divergence Index using population data from
the 2010 decennial census aggregated at the level of census
tracts\footnote{Census tracts are geographic units defined by the Census
  Bureau. They have an average population of 4,000 individuals and are
  intended to approximate neighborhoods. Most studies of residential
  segregation use census tract data. In cases where a tract is bisected
  by the city boundary, I create separate population counts for the city
  and suburban portions of the tract by aggregating census block-level
  data.} \citep{CensusSummary:bZv73ozJ}. I then decompose overall
segregation in the Detroit metro area into the segregation occurring
between the city of Detroit and the remainder of the metro area (the
``suburbs''), and the segregation occurring among the tracts within each
these subareas. The between-subarea component of segregation measures
how surprising the racial composition of each subarea is given the metro
area's overall racial composition. The within-subarea component of
segregation measures how surprising the racial composition of tracts
within each subarea is given each subarea's overall racial composition.
Total segregation for the metro area is the sum of the between-subarea
segregation and the average within-subarea segregation. The total is
equal to measuring the segregation of all tracts in the metro
area.\footnote{Note that it is not possible to use the Dissimilarity
  Index for this decomposition because it is not additively
  decomposable.} In the same fashion, I decompose relative homogeneity
into between- and within-subarea components with the Information Theory
Index.

The first key difference between the decompositions of the Divergence
Index and Information Theory Index is in the weighting of the
within-subarea values. Table \ref{tab:decompvals} demonstrates how the
decomposition formulas are applied in the Detroit metro area
decomposition. It reports the index values for each subarea, the weights
applied to each, and the additive value in the decomposition. The index
values for the individual subareas are weighted to obtain the additive
value, which is the subarea's contribution to the overall index.I also
present the percentage contribution of each component to the overall
index in Table \ref{tab:decomptable}. I present both Tables
\ref{tab:decompvals} and \ref{tab:decomptable} to provide greater
transparency in how the decomposition formulas are applied to the
empirical data and how the values for the percentage contributions are
obtained. I will start by describing the within-subarea component of the
decomposition, followed by the between-subarea component.

The additive value of the within-subarea component of the Divergence
Index is a population-weighted sum of the within-subarea index values
for all subareas (see Equation \ref{eq:D_decomp}). By applying
population weights, the index is able to represent the average
segregation experienced by individuals. The additive value of the
within-subarea components of \(H\) and \(D/E\) are population \emph{and}
relative diversity weighted sums of the within-subarea index values for
all subareas (see Equation \ref{eq:H_decomp}). Due to the additional
relative diversity weight, subareas that are more diverse contribute
more to the overall index than less diverse subareas.

An implication of this is that the way each subarea's index value
contributes to the within-subarea component and the overall metro index
differs for \(D\) and for \(H\) and \(D/E\). For \(D\), the within-city
weight is 0.17, and the within-suburbs weight is 0.83, which represents
their relative shares of the metro population. For \(H\) and \(D/E\),
the total weight is 0.09 for the within-city component and 0.54 for the
within-suburb component, which places more relative weight on the
within-suburb component than the within-city component, compared to
\(D\).

\vspace{1ex}
\centerline{[Table \ref{tab:decompvals} about here.]}

The second difference in the decomposition of the indexes concerns the
calculation of local values, as seen in the between-subareas component.
Table \ref{tab:decomptable} reports the percentage contribution of each
decomposition component to the overall indexes. The percentages are
calculated using the the additive values -- the component values after
the relevant weights have been applied. The indexes show the same
overall pattern, with the between-subareas component accounting for
about two-thirds of the metro's overall index value. The decomposition
reveals that the largest differences in both population composition and
diversity occur at the regional level --- \emph{between} Detroit and the
suburbs. There are comparatively less differences among the tracts
\emph{within} each subarea. However, if we take a closer look at the
local values of the between-subarea component in Table
\ref{tab:decomptable}, there is a stark difference between the indexes.
Results for \(D\) and \(D/E\) show that Detroit contributes more to the
between-subarea value than the suburbs, while results for \(H\) show the
opposite pattern.

\begin{figure}[h]
  \includegraphics[width=\linewidth, center]{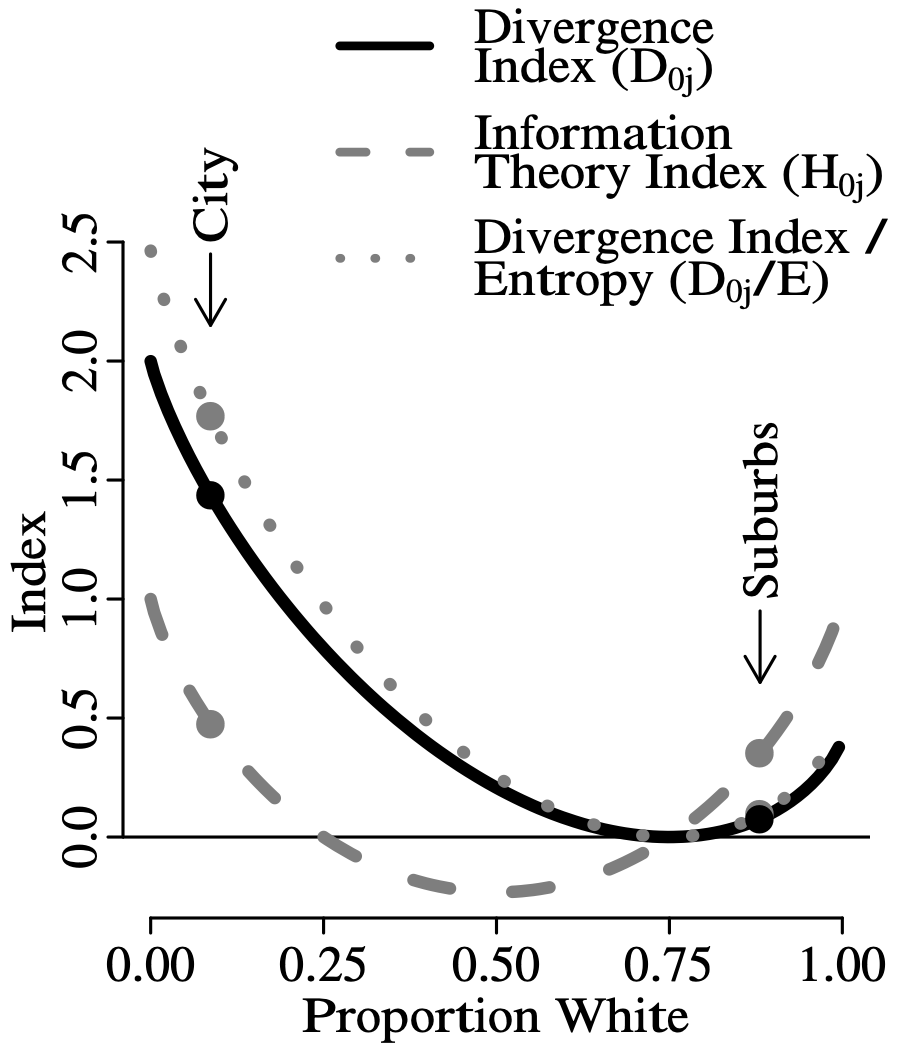}
  \captionsetup{width=\linewidth}
  \caption{Black-White Segregation and Diversity Between Detroit and the Suburbs}
  \label{fig:decompfig}
\end{figure}

Figure \ref{fig:decompfig} shows the functional form of the indexes for
the Detroit metro area population and indicates the local values of the
city and suburb components of the between-subarea index. The figure
shows the unweighted between-subarea index values --- the same values
found in the ``between-subareas'' rows and ``index value'' columns of
Table \ref{tab:decompvals}. The horizontal axis shows the proportion
White, and the vertical axis shows the index values. The solid line
shows the functional form of segregation measured with \(D_{0j}\), and
the dashed line shows relative homogeneity measured with \(H_{0j}\), and
the dotted line shows the value of \(D_{0j}/E\). The points in each
figure indicate the unweighted index value for each subarea --- the city
of Detroit and the suburbs. The figure shows the pronounced difference
between the between-subarea index values for the city and suburbs
measured with \(D\) and \(D/E\), but not with \(H\), which is similar
for both the city and suburbs.

The between-subarea Divergence Index measures the difference between
each subarea's composition and overall metro area composition. The
proportion White is 0.75 in the metro area, compared to 0.09 in Detroit
and 0.88 in the suburbs. From the perspective of the Divergence Index,
0.09 is a very surprising local proportion, more so than 0.88, given
that the overall proportion White is 0.75. Therefore, there is greater
divergence between the population compositions of Detroit and the metro
area than between the suburbs and the metro area, and greater divergence
indicates higher segregation. Detroit's between-subarea segregation is
sufficiently higher than the suburbs that even after weighting each
subarea's value by its share of the metro population (0.17 for Detroit
and 0.83 for the suburbs) Detroit's contribution to between-subarea
segregation is still larger than the suburbs's contribution. The pattern
is similar for the \(D/E\) measure.

Results for the Information Theory Index show an opposite trend: Detroit
contributes less to overall segregation than the suburbs. White
residents are over-represented in the suburbs and Black residents are
over-represented in Detroit, relative their metro proportions. But the
city and suburban populations both have about the same level of
diversity, and each has less diversity than the overall metro
population. The Information Theory Index is concerned only with the mix
of groups in each subarea relative to the metro, not the specific group
proportions. The difference in the percentage contributions of the city
and suburbs is mainly due to their share of the metro population, which
is much smaller for the city than the suburbs.

This analysis demonstrates key differences in the calculation and
interpretation of decomposition results for the Divergence Index and
Information Theory Index. First, the within-subarea components of each
index differ in the weights that are applied to each subarea. Both
indexes apply population weights, but \(H\) (and \(D/E\)) applies an
additional relative diversity weight, which means that more diverse
subareas contribute more to the within component and overall index than
less diverse subareas. Second, the between-subareas component calls our
attention to differences in the calculation of local values and what
each index is measuring. \(H\) measures the level of diversity in the
city and suburban populations relative to the overall diversity of the
metro area's population, and \(D\) measures the over/under
representation of groups in the city and suburban populations relative
to the metro area's population composition.

\vspace{1ex}
\centerline{[Table \ref{tab:decomptable} about here.]}

\vspace{1ex}
\centerline{[Figure \ref{fig:decompfig} about here.]}

\hypertarget{analyzing-changes-in-the-100-largest-metros-from-1990-to-2010}{%
\section{Analyzing Changes in the 100 Largest Metros from 1990 to
2010}\label{analyzing-changes-in-the-100-largest-metros-from-1990-to-2010}}

\label{R1_2b} In this section, I analyze and compare the Divergence
Index and Information Theory Index for Asian, Black, Latinx, and White
ethnoracial groups from 1990 to 2010 in the 100 most populous U.S.
metropolitan area divisions as of 2010.\footnote{Using the U.S. Census
  Bureau's categories of race and ethnicity, I define four mutually
  exclusive ethnoracial groups: Hispanic or Latino of any race
  (``Latinx''), non-Hispanic Asian and Pacific Islander (``Asian''),
  non-Hispanic Black (``Black''), and non-Hispanic White (``White'').
  For continuity across census years, the ``Asian'' ethnoracial group
  includes those who identified as Pacific Islanders. I use the
  metropolitan statistical area definitions that were updated by the
  Office of Management and Budget in 2013
  \citep{U.S.CensusBureau2013df}.} The indexes reveal different trends:
the Information Theory Index has tended to decrease overtime,
particularly within cities, while the Divergence Index has tended to
increase, particularly within suburbs. The different properties of the
indexes and how they respond to changes in local and overall population
composition offer insight into why we observe empirical differences
between the indexes.

I use U.S. Census data at the block level for each decade
\citep{CensusSummary:bZv73ozJ, NHGIS:2021}. I apply the 2010 boundaries
for blocks, central cities, and metropolitan areas to data from 1990 and
2000 to create constant area boundaries across time \citep{NHGIS:2021}.
Using stable areal units means that the changes that we observe in the
index values are net of the effect of any boundary changes that may have
occurred.

\begin{figure}[h]
  \includegraphics[width=\linewidth, center]{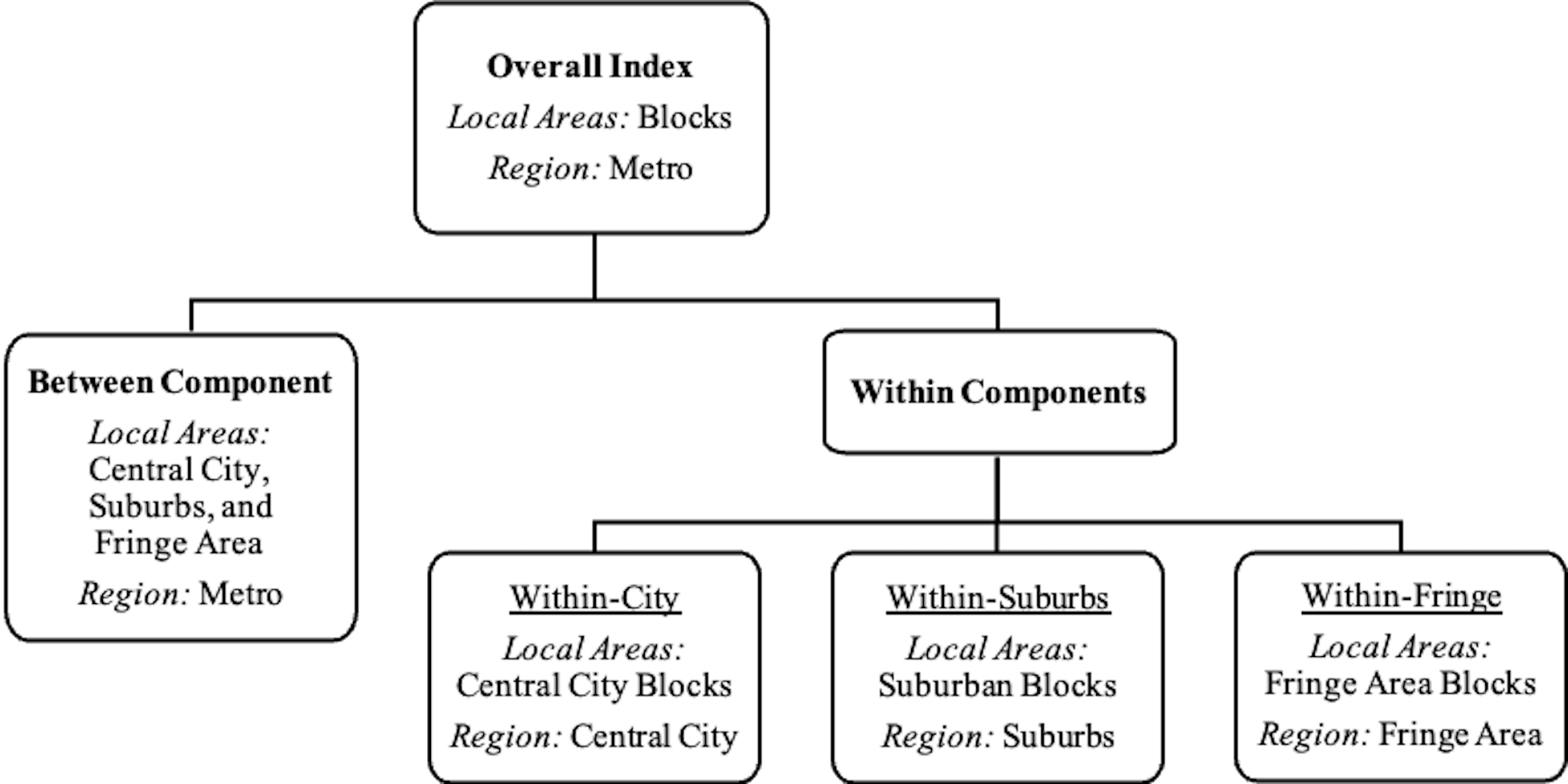}
  \captionsetup{width=\linewidth}
  \caption{Components of the Decomposition Within and Between Subareas of each Metropolitan Area}
  \label{fig:decfig}
\end{figure}

I decompose the overall indexes into within and between components,
using the primary central city, suburbs (census defined places), and
fringe (nonplace areas) of each metro as the subareas
\citep[see:][]{Lichter:2015gz, Parisi:2011wk, Owens2019bi}. Figure
\ref{fig:decfig} illustrates the relationship between the components of
the decomposition, with the between component on the left and the within
component and its three subcomponents on the right.

\vspace{1ex}
\centerline{[Figure \ref{fig:decfig} about here.]}

The decomposed index values and change over time are presented in Table
\ref{tab:chngHD}. For comparability, I have included the unweighted
index values for the within central city, within suburbs, and within
fringe areas subcomponents, rather than the weighted additive values,
because \(D\) and \(H\) apply different weights to each subcomponent:
\(D\) applies population weights and \(H\) applies population and
relative diversity weights when aggregating each subarea's value into
the within component and overall index. (The weights are provided in
Equations \ref{eq:H_decomp} and \ref{eq:D_decomp}, and the discussion of
Table \ref{tab:decompvals} explains the implications of the different
weighting.)

\vspace{1ex}
\centerline{[Table \ref{tab:chngHD} about here.]}

Overall, the within component of the decomposition accounts for the
majority of the overall segregation and relative homogeneity in the
metro areas across all time periods. On average, the differences in
population composition and diversity between the central city, suburbs,
and fringe area of each metro is relatively small compared to the
differences among the blocks within each subarea. This is consistent
with the findings of previous decomposition studies using the
Information Theory Index
\citep{Owens2019bi, Lichter:2015gz, Farrell.2014}.

\hypertarget{change-overtime}{%
\subsection{Change Overtime}\label{change-overtime}}

The \(H\) index reveals a dominant trend of declines in overall metro
\(H\) from 1990 to 2010 in 93 of the 100 metros, which is consistent
with the findings of Lichter, Parisi, and Taquino
\citeyearpar{Lichter:2015gz} and other studies
\citep[e.g.,][]{Elbers.2021, Parisi:2015ci, Kye2023ir}.\footnote{Note
  that \citet{Lichter:2015gz} use different race group combinations,
  include 222 metropolitan areas, and further decompose the
  within-suburbs component in their decomposition. Nonetheless, their
  results and those shown here have notable similarities.} The average
change across metros is a decrease of 0.081. (See Table
\ref{tab:chngHD}.) Also consistent with previous studies, the decreases
were more pronounced for the within component of \(H\), compared to the
between component.

Changes in overall metro \(D\) over this period show a different trend,
with decreases in only 27 of the metros. The average change across the
100 metros is an increase of 0.036. The largest increases were in the
within suburb and within fringe area segregation (0.056 each). In
contrast, the segregation within central cities decreased on average,
though to a lesser degree than the decreases in \(H\).

Comparing patterns of change for individual metros in the overall metro
indexes and the within components of the indexes reveals opposite trends
in two-thirds of the metros: \(H\) decreases and \(D\) increases from
1990 to 2010. (See Table \ref{tab:HD2by2}.) The contrast between the
indexes is less pronounced for the between component: only one-quarter
of the metros show opposite trends for \(H\) and \(D\).

To summarize, the overall and within component of the \(H\) index
\emph{decrease} in all but a few metros, and the between component of
the index decreases in about two-thirds of the metros. The overall and
within component of the \(D\) index \emph{increase} in three-quarters of
the metros, and the between component of segregation increases in about
two-thirds of the metros of the metros.

\vspace{1ex}
\centerline{[Table \ref{tab:HD2by2} about here.]}

\hypertarget{within-city-and-within-suburbs-changes-overtime}{%
\subsection{Within-City and Within-Suburbs Changes
Overtime}\label{within-city-and-within-suburbs-changes-overtime}}

\begin{figure}[h]
  \begin{subfigure}{\linewidth}
    \includegraphics[width=0.75\linewidth, center]{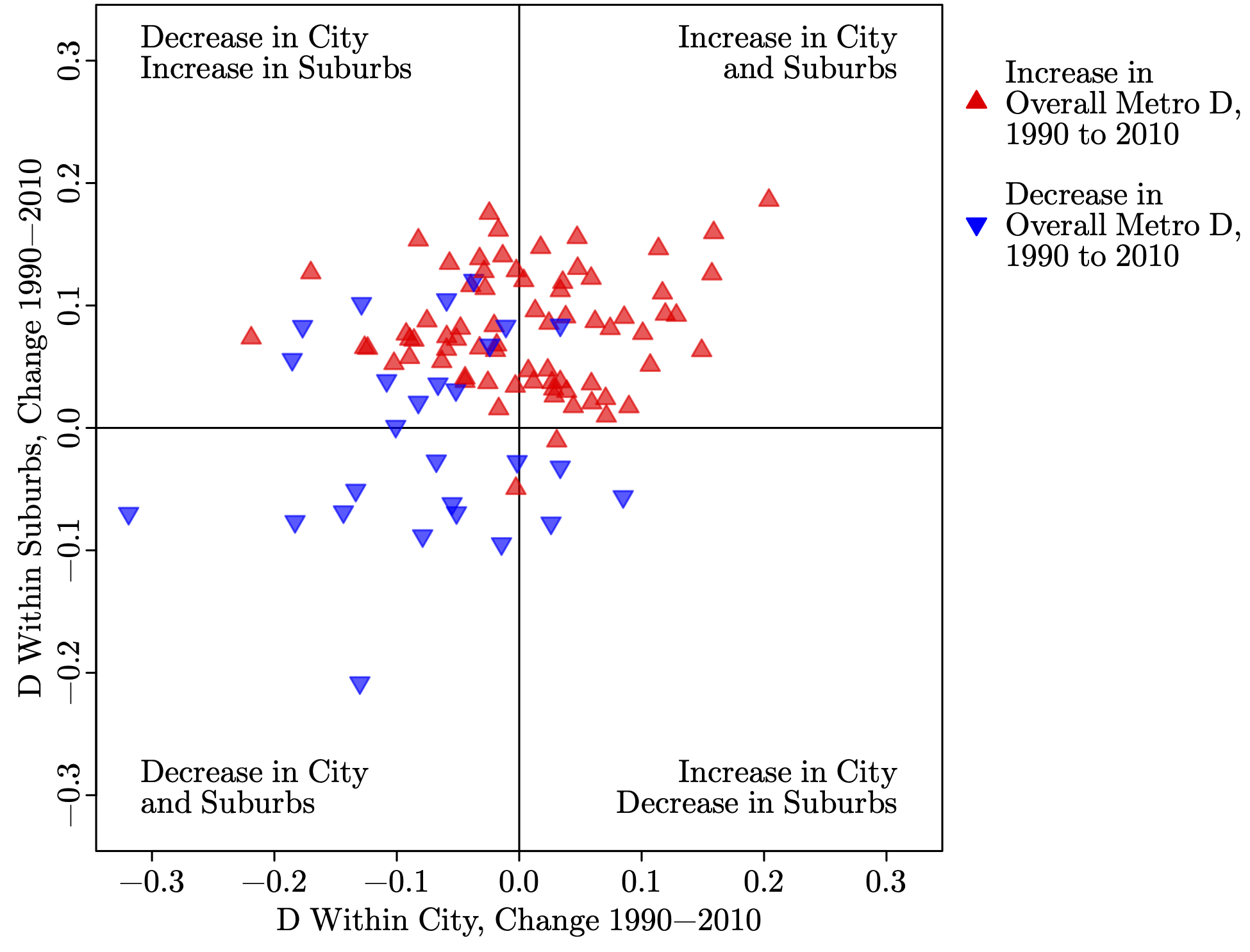}
    \caption{Divergence Index}
    \label{fig:withinD}
\vspace{3ex}
\end{subfigure}  
  \begin{subfigure}{\linewidth}
    \includegraphics[width=0.75\linewidth, center]{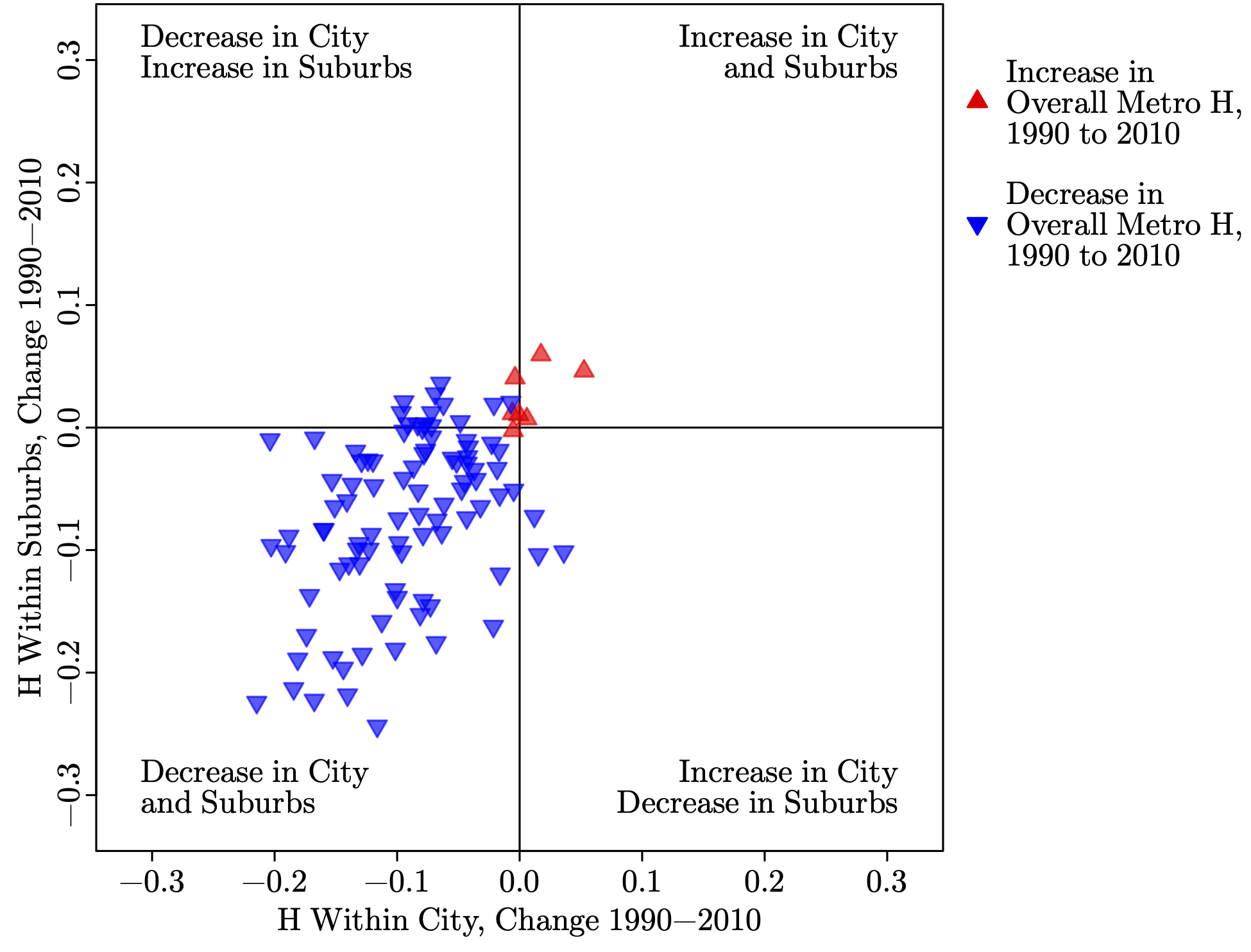}
    \caption{Information Theory Index}
    \label{fig:withinH}
  \end{subfigure}  
  \captionsetup{width=\linewidth}
  \caption{Changes in Within-City and Within-Suburb Index Values for the Divergence Index and Information Theory Index in the 100 Largest Metropolitan Areas, 1990 to 2010}
  \label{fig:withinHD}
  \begin{minipage}{\linewidth}
  \vspace{2ex}
  \small Data Source: Author's calculations of block-level population data from the decennial census using 2010 boundaries \citep{CensusSummary:bZv73ozJ, NHGIS:2021}
  \end{minipage}
\end{figure}

Analyzing the within-city and within-suburbs subcomponents of the
indexes for each metro area provides a further comparison of changes
from 1990 to 2010. Figure \ref{fig:withinHD} shows the relationship
between the within-city and within-suburb changes over this time period
for the Divergence Index (Figure \ref{fig:withinD}) and Information
Theory Index (Figure \ref{fig:withinH}). The four quadrants of each
panel indicate metros where there were increases in both the within-city
and within-suburb measures (upper right), increases in within-city and
decreases in within-suburb measures (lower right), decreases in both
within-city and within-suburb measures (lower left), decreases in
within-city and increases in within-suburb measures (upper left). The
color and shape of the points indicates the trend of the overall metro
measures --- the blue down-pointing triangle indicates decreases in the
metro index and the red up-pointing triangle indicates increases in the
metro index from 1990 to 2010 --- and illustrate the finding of
decreasing metro \(H\) and increasing metro \(D\) in most metro areas
over this period.

\vspace{1ex}
\centerline{[Figure \ref{fig:withinHD} about here.]}

There are two key takeaways from Figure \ref{fig:withinHD}. First,
consistent with the overall metro \(D\) trend, most of the metros (84 of
100) show increases in within-suburb \(D\), as indicated by the majority
of the points falling in the top two quadrants of Figure
\ref{fig:withinD}. In the suburbs of these metro areas, the ethnoracial
composition of blocks has become increasingly different from the overall
suburban population composition over time. These trends are found amid
overall increases in diversity (measured with the Entropy Index) at each
level of aggregation: the metros, cities, suburbs, and fringe areas.
(See Table \ref{tab:metroE} in Appendix E.) In fact, the biggest
increases in diversity occurred within suburbs.

However, many of these same metros (46) show accompanying decreases in
within-city \(D\), as indicated by the large number of points falling in
the top left quadrant of Figure \ref{fig:withinD}. These results are
consistent with the literature that finds declines in neighborhood-level
segregation within cities alongside the increasing prevalence of
``ethnoburbs'' \citep{Li1998an} in suburban places
\citep{Lee2014er, Logan:2010gs, Lichter:2013et}.

Second, nearly all metros (94 of 100) show decreases in within-city
\(H\), as indicated by the majority of the points falling in the left
two quadrants of Figure \ref{fig:withinH}. In these metro areas, the
racial diversity of city blocks has come to more closely resemble the
overall diversity of the city population. In most of these metros (78),
within-suburb \(H\) has also decreased, as indicated by the large number
of points falling in the bottom left quadrant of Figure
\ref{fig:withinH}. There were relatively small increases in
within-suburb \(H\) in 19 of the metros, shown in the upper left
quadrant of Figure \ref{fig:withinH}, suggesting that the diversity of
blocks within these suburban communities did not quite keep pace with
the rising racial diversity of the suburbs as a whole during this time
period.

\hypertarget{differences-between-the-indexes}{%
\subsection{Differences between the
Indexes}\label{differences-between-the-indexes}}

The different properties of \(D\) and \(H\) and how they respond to
changes in local and overall population composition offer insight into
why we observe empirical differences between the indexes.\\
When there are equal proportions of each group, the \(H\) and \(D\)
indexes are the same. The difference between the two measures increases
as the overall group proportions become more unequal. Practically what
this means is that \(D\) adjusts the expected local proportions of an
even distribution as the overall population composition changes. Whereas
\(H\) adjusts the expected local mix of groups as the overall diversity
changes.

For example, in a metro area with one very large group and one small
group, like city C in Figure \ref{fig:cityC}, there is less potential
for high levels of \(D\) than in a second metro area where the groups
are about equal in size, like city A in Figure \ref{fig:cityA}. If group
members do not co-reside in any local areas, each of the metros would be
maximally segregated.

In the first metro, local areas composed entirely of the group that is a
small proportion of the metro population would be the most surprising
--- their composition is very different than the expected local
proportion of each group if there is no segregation. These areas would
have the highest local segregation values, but the number of such areas
would be limited given the overall population composition. Local areas
composed entirely of the group that is a very large proportion of the
metro population would not be too different from the expected
proportions --- the metro's population composition --- and would
contribute little to overall segregation.

In the second metro area, with groups that are about equal in size,
\(D\) expects groups to be equally represented in local areas if there
is no segregation in the metro. It would be very surprising to observe
local areas composed entirely of one group --- more surprising than a
local area in the first metro composed entirely of the very large group
--- and this could occur in every local area of the metro, generating
very high levels of segregation.

In contrast with \(H\) any local area in the first metro with the same
relative group proportions --- regardless of which group is large and
which is small --- would have a zero index value. Local areas with more
equal group sizes than the metro would have a negative index value. Any
local area composed entirely of one group would have a positive index
value of 1, regardless of whether that group is large or small in the
overall metro population. In the second metro, results would be similar
for \(H\) and \(D\), since the indexes would return the same value when
there are equally sized groups.

\hypertarget{a-note-about-composition-invariance.}{%
\subsubsection{A Note about Composition
Invariance.}\label{a-note-about-composition-invariance.}}

It may seem counter-intuitive that there could be lower segregation in
the first metro area than the second: with one very large group in the
first metro, there are likely to be local areas composed of that group
alone, which would seem to indicate very high segregation. However,
measures of over/under representation necessarily adjust for the
expectation of an even or random local distribution of the group
populations using the overall proportion of each group in the region as
the baseline for comparison.

There has been a long debate over whether segregation indexes should be
compositionally invariant, or free from margin dependence. In recent
years, methods have been developed to isolate differences in segregation
that are due to differences in the distribution of the population across
areal units and groups (also called unit and group marginals), and
``structural change'' (also called ``pure segregation'')
\citep[e.g.,][]{Elbers2023ms, Mora:2011ii, Mora2009ip}. However, a key
limitation of compositionally invariant indexes or focusing only on
``structural change'' is the assumption that differences in the marginal
distributions and differences in the ``structural'' component are
independent of one another. There is a wealth of segregation literature
that suggests that changes in the marginals, especially group
proportions, may activate mechanisms associated with segregation, such
as, racially restrictive covenants \citep[e.g.,][]{Rothstein2017cl},
exclusionary zoning \citep[e.g.,][]{Rothwell2009ed}, White flight
\citep[e.g.,][]{Lichter:2015gz}, and racial steering by real estate
agents \citep[e.g.,][]{Besbris:2017hn}. A poignant example is the Great
Migration, in which millions of Black people moved from the South to
Northern, Midwestern, and Western states between the 1910s and 1970s.
Racist policies and practices in response to these demographic changes
fueled housing discrimination against Black southerners and created
segregated neighborhoods.

The Divergence Index is not compositionally invariant by design. The
index uses the overall population composition as the comparative
reference for segregation, which captures our expectation about what the
local compositions would be if there is no segregation. If there is an
influx of a population group to a city from one time period to the next,
the comparative reference adapts to represent our updated expectation
about the local compositions: the compositions of all local areas should
reflect this influx if there is no segregation in the city.
Compositional changes from demographic processes, including migration in
and out of the region and residential resorting within the region, are
relevant for understanding segregation patterns and change. Making the
index invariant to differences in population composition, would create a
substantively different index that no longer carries the same meaning or
interpretation. (For a longer discussion of the topic of compositional
invariance, see Appendix B.)

\hypertarget{how-the-indexes-respond-to-changes-in-diversity.}{%
\subsubsection{How the Indexes Respond to Changes in
Diversity.}\label{how-the-indexes-respond-to-changes-in-diversity.}}

As metro areas became more diverse from 1990 to 2010, the \(D\) index
adjusts the expected local proportions of an even distribution
accordingly. Drawing on the previous example, if the group proportions
of a metro changed from resembling the first metro area --- with one
very large group and one small group --- in 1990 to looking more like
the second metro area --- with similarly sized groups --- by 2010, then
blocks composed entirely of one ethnoracial group would be more
segregated in 2010 than they were in 1990, and segregation in the metro
area and subareas would likely increase. In order for metro segregation
to remain the same or decrease over time, current residents would need
to relocate from blocks where their ethnoracial group is overrepresented
to blocks where their ethnoracial group is underrepresented, newcomers
to the metro area would need to move to blocks where their ethnoracial
group is underrepresented, or current residents of blocks where their
ethnoracial group is overrepresented would need to move out of the metro
area. The observed increases in the \(D\) index for metros from 1990 to
2010 suggest that population redistribution has not occurred in this way
or has not kept pace with changes to the overall metro composition.

By comparison, \(H\) compares local diversity in each block to overall
diversity in the metro. \(H\) adjusts the relative mix of groups
expected in each block as the overall metro diversity changes. If the
group proportions of a metro changed from resembling the first metro
area in 1990 to looking more like the second, more diverse metro area by
2010, the index value for a block composed entirely of one ethnoracial
group would not increase, as it does with \(D\). Instead, it would have
the same index value in 2010 as it did in 1990, a value of 1.

In order for the \(H\) index to remain the same or decrease over time,
current residents would need to relocate from blocks where the
ethnoracial group proportions are unequal and their group is the largest
to another block with unequal proportions and their group is the
smallest, newcomers to the metro area would need to move to blocks with
unequal proportions and their group is the smallest, or current
residents of blocks with unequal proportions and their group is the
largest would need to move out of the metro area.

Some of the moves that would lead to decreases in \(H\) would be
associated with increases in \(D\). For example, in the first metro area
with one large group and one small group, if members of the largest
group move from blocks where the group proportions match the metro
composition to blocks where the group proportions are inverted, so that
the largest group in the metro is the smallest group in the block and
the smallest group in the metro is the largest in the block, the \(H\)
index would decrease and the \(D\) index would increase in those blocks.

The empirical results suggests that \(H\) is capturing the tendency for
blocks to increase in diversity at roughly the same pace as the region,
but \(D\) is capturing the tendency for diversity promoting moves to
reproduce or intensify segregation.

\hypertarget{changes-in-segregation-and-diversity-within-suburbs}{%
\subsection{Changes in Segregation and Diversity within
Suburbs}\label{changes-in-segregation-and-diversity-within-suburbs}}

\begin{figure}[h]
  \includegraphics[width=\linewidth, center]{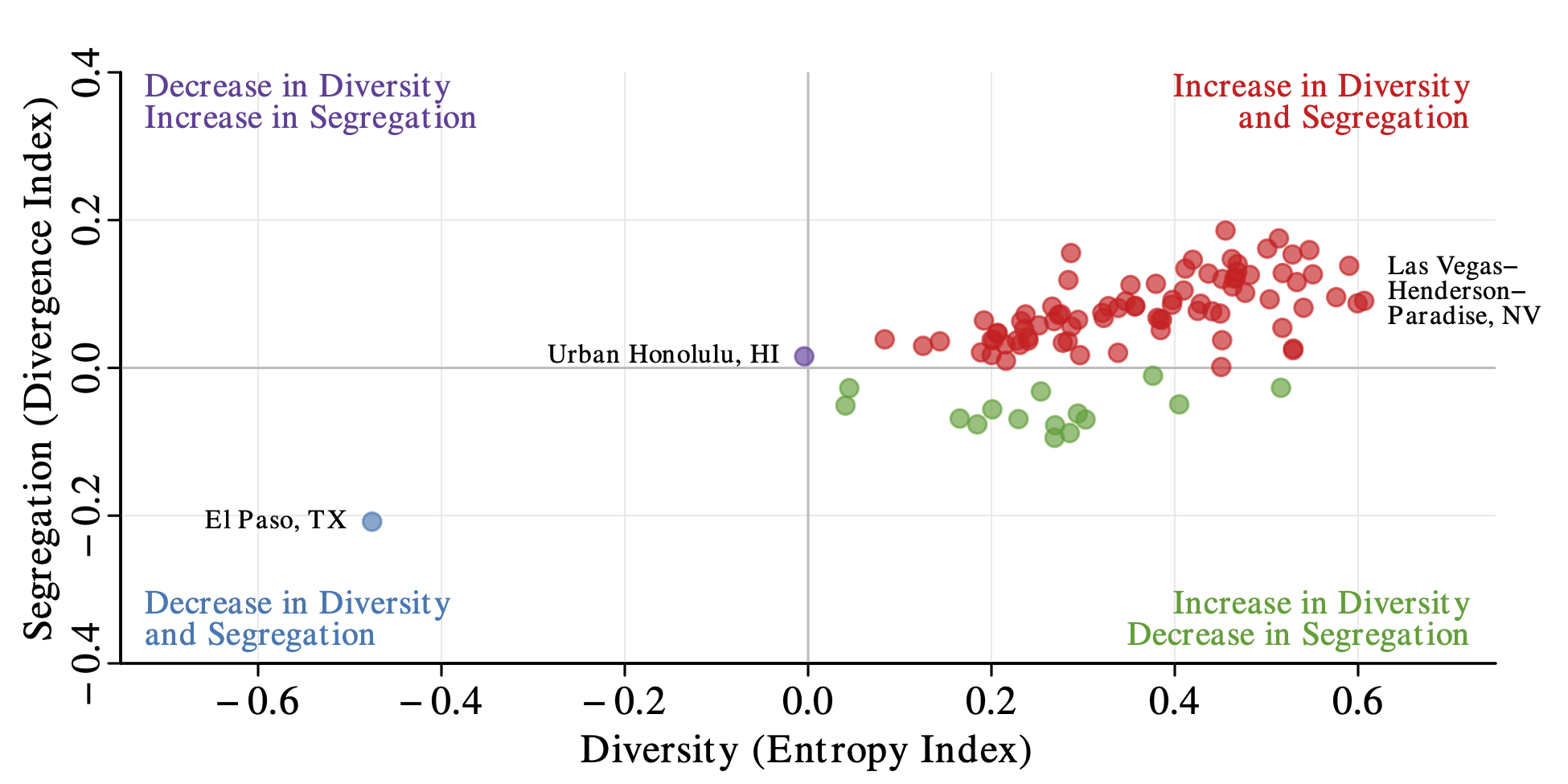}
  \captionsetup{width=\linewidth}
  \caption{Changes in Within-Suburb Segregation and Diversity in the 100 Largest Metropolitan Areas, 1990 to 2010}
  \label{fig:winsubDE}
  \begin{minipage}{\linewidth}
  \vspace{2ex}
  \small Data Source: Author's calculations of block-level population data from the decennial census using 2010 boundaries \citep{CensusSummary:bZv73ozJ, NHGIS:2021}
  \end{minipage}
\end{figure}

An informative approach for understanding the dimensions and dynamics of
residential differentiation is to analyze segregation alongside
diversity using the Divergence and Entropy Indexes. As an illustration,
Figure \ref{fig:winsubDE} shows the relationship between changes in
within-suburb diversity and segregation from 1990 to 2010. The four
quadrants of the figure indicate the suburbs of metros where there were
increases in both the segregation and diversity (upper right, red),
increases in diversity and decreases in segregation (lower right,
green), decreases in both segregation and diversity (lower left, blue),
decreases in diversity and increases in segregation (upper left,
purple).

As noted earlier, both segregation and diversity within suburbs tended
to increase over this period, as demonstrated by the heavy concentration
in the upper right quadrant of Figure \ref{fig:winsubDE}. Although there
is a tendency for segregation and diversity to increase together, the
relationship is not deterministic. Fifteen of the metros whose suburbs
increased in diversity experienced decreases in segregation.

\vspace{1ex}
\centerline{[Figure \ref{fig:winsubDE} about here.]}

As an example, the suburbs of the Jackson, MS, and Columbia, SC metros
had similar increases in diversity and levels of diversity in 1990 and
2010 (see Table \ref{tab:winsubDE_eg}). The suburbs of both metros also
had similar population sizes and racial demographics in each time
period. In 1990, the Jackson suburbs had considerably more segregation
than the Columbia suburbs (0.60 and 0.46). But by 2010, their levels of
segregation were nearly the same (0.53 and 0.54). This change represents
a decrease in segregation in the Jackson suburbs and an increase in
segregation in the Columbia suburbs, despite both experiencing similar
increases in diversity.

As this example illustrates, the average trends in segregation and
diversity do not imply a deterministic relationship. The variation in
the direction and strength of the relationship between segregation and
diversity suggest different underlying population processes across metro
areas. For example, in some places increases in diversity may be
associated with stable residential integration, while in others the
increases may be driven by processes of White flight and racial
turnover: White residents may leave predominantly White, segregated
neighborhoods as new non-White residents enter the neighborhood,
generating fleeting increases in diversity as the process unfolds and
reproducing segregation in the long run
\citep{Krysan2017cs, Kye2023ir, Farrell:2011ix, Rastogi2020ce}.

\vspace{1ex}
\centerline{[Table \ref{tab:winsubDE_eg} about here.]}

\hypertarget{conclusion}{%
\section{Conclusion}\label{conclusion}}

In this paper, I have highlighted the conceptual, mathematical, and
empirical distinctions between segregation and diversity and introduced
the Divergence Index as a decomposable measure of segregation.
\label{R1_sp8} Decomposition analysis is a critical tool for examining
the social and spatial dimensions of diversity, segregation, and
inequality. Previous scholarship has used the Information Theory Index
as the best alternative to the Dissimilarity Index to decompose
segregation within and between communities, municipalities, or school
districts
\citep[e.g.,][]{Bischoff:2008bw, Farrell:2008hh, Fischer:2008kx, Fischer:2004tk, Parisi:2011wk, Fowler:2016bk, Fiel:2013ja, Owens2019bi, Owens2023lb, Lichter:2015gz},
however the indexes represent different concepts. Although studies have
acknowledged this difference, they commonly interpret the Information
Theory Index as carrying the same meaning as the Dissimilarity Index.
This blurs the conceptual distinction between segregation and diversity,
as well as having empirical implications for our understandings of
segregation, diversity, and their relationship.

I illustrated what can be learned through decomposition analyses with
the case of Westchester County, which highlights the implications of
decisions about the placement of affordable housing on reducing (or
increasing) segregation in the county. Next, I provided a brief review
of entropy-based measures of diversity before introducing the Divergence
Index, and I compared the mechanics of the Divergence Index and
Information Theory Index using stylized and empirical examples.

I illustrated the importance of preserving the conceptual distinction
between segregation and diversity with an empirical analysis. I
decomposed and compared the Divergence Index and Information Theory
Index for Asian, Black, Latinx, and White ethnoracial groups from 1990
to 2010 in the 100 most populous U.S. metropolitan areas. The indexes
reveal different trends: the Information Theory Index has tended to
decrease overtime, particularly within cities, while the Divergence
Index has tended to increase, particularly within suburbs.

I offered insight into why we observe empirical differences between the
indexes by explaining the different properties of indexes and how they
respond to changes in local and overall population composition. I also
highlighted that despite the general trend of increases in segregation
alongside increases in diversity, the relationship is not deterministic.
Differences across metros in the direction and strength of the
relationship suggests the possibility of different underlying
mechanisms, where in some metros increases in diversity may reflect the
transience of racial turnover that eventually reproduces segregation,
while in others it may be associated with the gradual formation of
stably integrated communities.

There is no one perfect measure. In this paper, I have highlighted the
distinctions between multiple indexes to promote a greater understanding
of what each index is measuring and their potential use cases, and to
help scholars make informed choices about which index(es) are most
appropriate for a given analysis.

Segregation and diversity are important aspects of residential
differentiation. Rather than being a substitute for measures of
diversity, the Divergence Index is an additional tool for measuring and
decomposing segregation. It complements existing measures by enabling
the analysis and decomposition of segregation alongside diversity. The
Divergence Index offers a distinct lens, which enables richer, deeper
understandings of the dimensions and dynamics of segregation.

\setlength{\leftskip}{\normleftskip}
\setlength{\parindent}{\normparindent}

\clearpage

\section{Endnotes}\label{endnotes}
\vspace{-10ex}{\parskip 2ex \theendnotes}

\let\footnote=\origfootnote
\renewcommand{\notesname}{}
\renewcommand{\footnotemark}{\origfootnotemark}
\renewcommand{\footnotetext}{\origfootnotetext}
\renewcommand{\thefootnote}{\roman{footnote}}

\clearpage

\hypertarget{references}{%
\section{References}\label{references}}

\vspace{-5ex}

\begingroup

\renewcommand{\refname}{}

\bibliographystyle{asr} 
  \bibliography{Bib_CurrentLibrary}

\endgroup

\clearpage

\hypertarget{tables}{%
\section{Tables}\label{tables}}

\begin{table}[h]
  \captionsetup{width=\linewidth}
  \caption{Hypothetical Neighborhood Group Proportions and Index Values in City B
  \label{tab:nbhdsegdiv}}
  \includegraphics[width=4in, center]{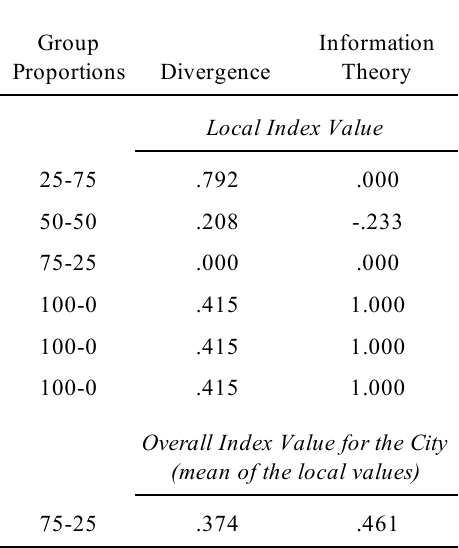}
  \vspace{10ex}
\end{table}

\begin{table}[h]
  \captionsetup{width=\linewidth}
  \caption{Decomposition of Black-White Indexes in the Detroit Metro Area
  \label{tab:decompvals}}
  \includegraphics[width=7in, center]{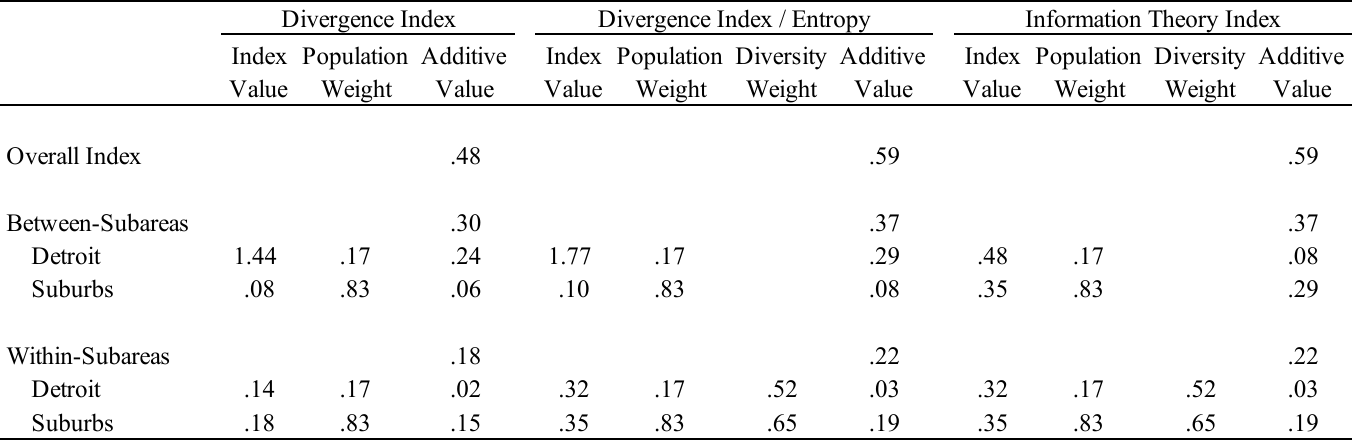}
  \begin{minipage}{\linewidth}
  \vspace{2ex}
  \small Data Source: Author's calculations of population data from the 2010 decennial census aggregated at the level of census tracts \citep{CensusSummary:bZv73ozJ}
  \end{minipage}
  \vspace{10ex}
\end{table}

\begin{table}[h]
  \captionsetup{width=\linewidth}
  \caption{Percentage Decomposition of Black-White Indexes in the Detroit Metro Area~(Percent of Overall Index)
  \label{tab:decomptable}}
  \includegraphics[width=4in, center]{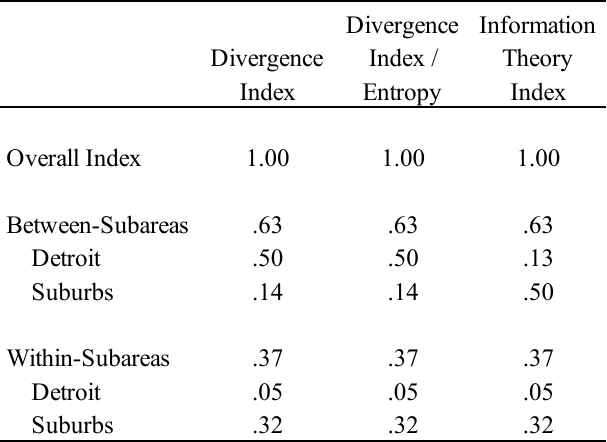}
  \begin{minipage}{\linewidth}
  \vspace{2ex}
  \small Data Source: Author's calculations of population data from the 2010 decennial census aggregated at the level of census tracts \citep{CensusSummary:bZv73ozJ}
  \end{minipage}
  \vspace{10ex}
\end{table}

\begin{table}[h]
  \captionsetup{width=\linewidth}
  \caption{Decomposition of Asian-Black-Latinx-White Divergence Index ($D$) and Information Theory Index ($H$) from 1990 to 2010, Mean for the 100 Largest Metropolitan Areas
  \label{tab:chngHD}}
  \includegraphics[width=7in, center]{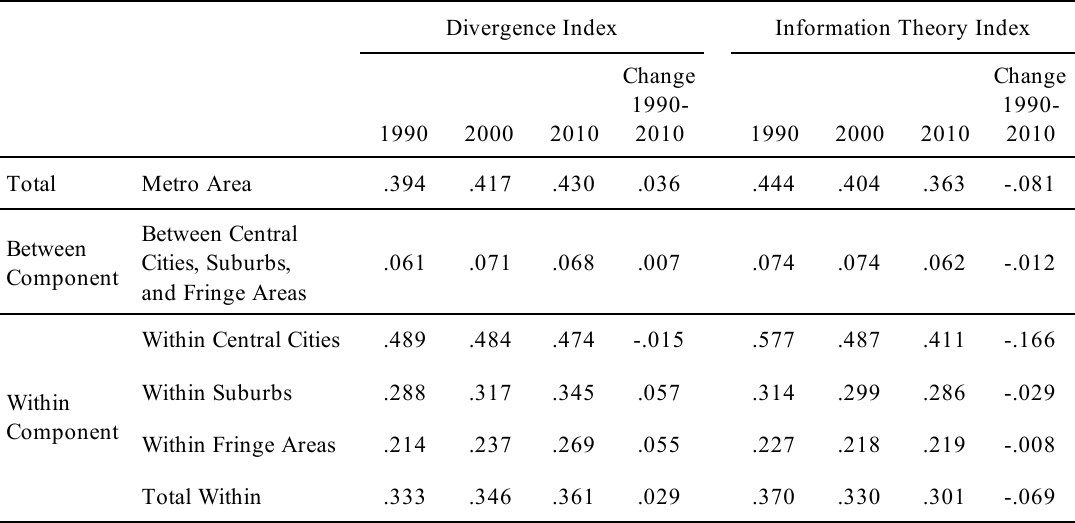}
  \begin{minipage}{\linewidth}
  \vspace{2ex}
  \small Data Source: Author's calculations of block-level population data from the decennial census using 2010 boundaries \citep{CensusSummary:bZv73ozJ, NHGIS:2021}
  \end{minipage}
\end{table}

\begin{table}[h]
  \captionsetup{width=\linewidth}
  \caption{Comparison of Changes for the Decomposed Components of the Asian-Black-Latinx-White Divergence Index ($D$) and Information Theory Index ($H$) from 1990 to 2010 in the 100 Largest Metropolitan Areas
  \label{tab:HD2by2}}
  \includegraphics[width=\linewidth, center]{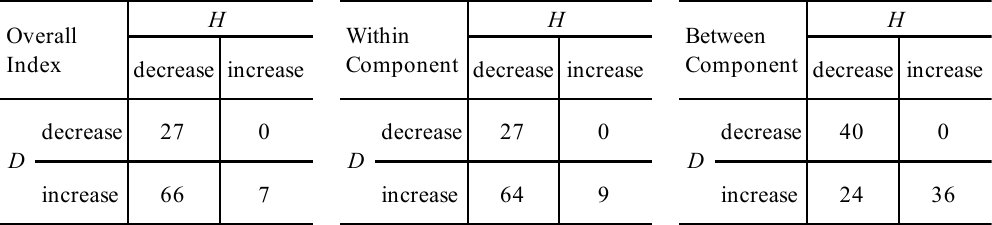}
  \begin{minipage}{\linewidth}
  \vspace{2ex}
  \small Data Source: Author's calculations of block-level population data from the decennial census using 2010 boundaries \citep{CensusSummary:bZv73ozJ, NHGIS:2021}
  \end{minipage}
  \vspace{10ex}
\end{table}

\begin{table}[h]
  \captionsetup{width=\linewidth}
  \caption{Asian-Black-Latinx-White Segregation ($D$) and Diversity ($E$) within the Suburbs of the Jackson, MS, and Columbia, SC, Metropolitan Areas from 1990 to 2010
  \label{tab:winsubDE_eg}}
  \includegraphics[width=\linewidth, center]{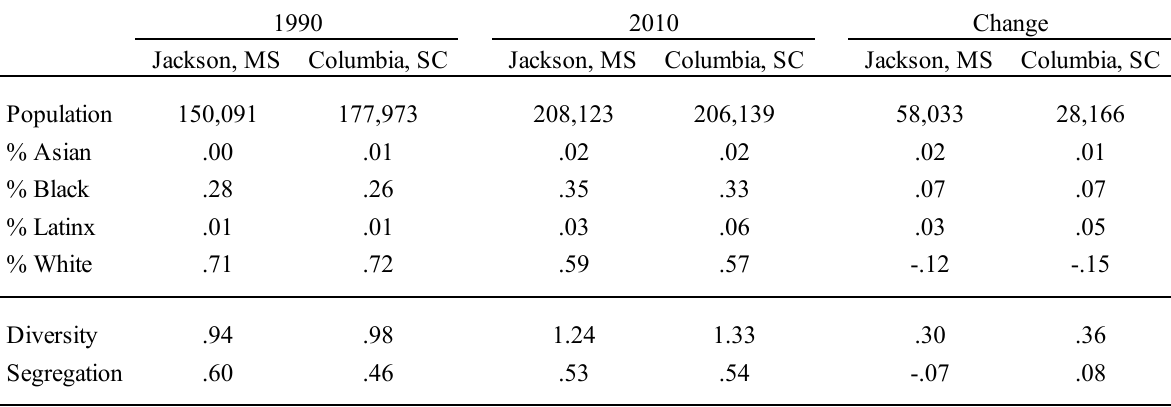}
  \begin{minipage}{\linewidth}
  \vspace{2ex}
  \small Data Source: Author's calculations of block-level population data from the decennial census using 2010 boundaries \citep{CensusSummary:bZv73ozJ, NHGIS:2021}
  \end{minipage}
  \vspace{10ex}
\end{table}

\clearpage

\hypertarget{figures}{%
\section{Figures}\label{figures}}

\vspace{10ex}

\vspace{10ex}

\clearpage

\appendix

\hypertarget{appendix-a}{%
\section{Appendix A}\label{appendix-a}}

\hypertarget{alternative-equation-for-the-overall-divergence-index}{%
\subsection{Alternative Equation for the Overall Divergence
Index}\label{alternative-equation-for-the-overall-divergence-index}}

At the aggregate level of a city or region, we can rewrite the equation
for the Divergence Index as: \(D=E-\bar{E_i}\). No such
reformulation exists at the local-level for locations or districts
within a city or region.

We use the following quantities in the equations below:
\(\tau_i\) is the population count for location \(i\), 
\(T\) is the overall population count, 
\(\pi_{im}\) is group \(m\)'s proportion of the population in location \(i\), and 
\(\pi_m\) is group \(m\)'s proportion of the overall population.  

\noindent The entropy for location \(i\) is: \begin{align*}
  E_i &=\sum_{m=1}^{M}{\pi_{im}\log{\cfrac{1}{\pi_{im}}}}
\end{align*} The average local entropy is the population-weighted mean
of the local entropies: \begin{align*}
\bar{E_i} &= \sum_{i=1}^{N}{\cfrac{\tau_i}{T}E_i} \\
          &= \sum_{i=1}^{N}{\cfrac{\tau_i}{T}\sum_{m=1}^{M}{\pi_{im}\log{\cfrac{1}{\pi_{im}}}}}
\end{align*} The region's entropy is: \begin{align*}
E &= \sum_{m=1}^{M}{\pi_{m}\log{\cfrac{1}{\pi_{m}}}}
\end{align*}

\noindent The Divergence Index for location \(i\) is: \begin{align*}
D_i &= \sum_{m=1}^{M}{\pi_{im}\log{\cfrac{\pi_{im}}{\pi_m}}}
\end{align*} The overall Divergence Index for the region is the
population-weighted mean of the local divergences: \begin{align*}
D &= \bar{D_i} = \sum _{i=1}^{N}{\frac{\tau_i}{T}D_i}
\end{align*}

\noindent Using these equations, we can rewrite the equation for \(D\) as: \begin{align*}
  D &= \sum_{i=1}^{N}{\frac{\tau_i}{T}\sum_{m=1}^{M}{\pi_{im}\log{\cfrac{\pi_{im}}{\pi_m}}}} \\
    &= \sum_{i=1}^{N}{\frac{\tau_i}{T}\sum_{m=1}^{M}{\pi_{im}\log{\cfrac{1}{\pi_m}}}} + \sum_{i=1}^{N}{\frac{\tau_i}{T}\sum_{m=1}^{M}{\pi_{im}\log{\pi_{im}}}} \\
    &= \sum_{m=1}^{M}{\pi_{m}\log{\cfrac{1}{\pi_m}}} - \sum_{i=1}^{N}{\frac{\tau_i}{T}\sum_{m=1}^{M}{\pi_{im}\log{\cfrac{1}{\pi_{im}}}}} \\
    &= E - \bar{E_i} \\
\end{align*}

\clearpage

\setcounter{table}{0}
\renewcommand{\thetable}{B\arabic{table}}

\hypertarget{appendix-b}{%
\section{Appendix B}\label{appendix-b}}

\hypertarget{desirable-properties-of-measures}{%
\subsection{Desirable Properties of
Measures}\label{desirable-properties-of-measures}}

Previous research has identified a set of desirable properties for
inequality and segregation measures
\citep{James:1985ti, Jahn:1947vj, Taeuber:1965us, Morgan:1981to, Coleman:1982vl, Allison:1978vw, Bourguignon:1979vj, Schwartz:1980tb, White:1986vx, Reardon:2002vt, Reardon:2004vl}.
Measures are commonly evaluated with respect to how well they meet these
criteria.

First, I review the criteria concerning the conceptual and
methodological qualities of measures. They address how measures should
respond to distributional changes (e.g., changes to the distribution of
individual incomes or the population count of each group). I organize
these criteria into three categories: features of the distribution,
changes to the whole distribution, and changes within the distribution.
Next, I review the desirable technical qualities and quantities of
measures. This second set of criteria address how a measure should be
calculated and interpreted.

\hypertarget{conceptual-and-methodological-qualities-of-measures.}{%
\subsubsection{Conceptual and Methodological Qualities of
Measures.}\label{conceptual-and-methodological-qualities-of-measures.}}

Measures should be invariant to the following features of a distribution
(Table \ref{tab:criteria1}):

\begin{longtable}[]{@{}
  >{\raggedright\arraybackslash}p{(\columnwidth - 4\tabcolsep) * \real{0.1512}}
  >{\raggedright\arraybackslash}p{(\columnwidth - 4\tabcolsep) * \real{0.4186}}
  >{\raggedright\arraybackslash}p{(\columnwidth - 4\tabcolsep) * \real{0.4302}}@{}}
\caption{Criteria Concerning Features of the Distribution
\label{tab:criteria1}}\tabularnewline
\toprule\noalign{}
\begin{minipage}[b]{\linewidth}\raggedright
Criteria
\end{minipage} & \begin{minipage}[b]{\linewidth}\raggedright
Description
\end{minipage} & \begin{minipage}[b]{\linewidth}\raggedright
Citations
\end{minipage} \\
\midrule\noalign{}
\endfirsthead
\toprule\noalign{}
\begin{minipage}[b]{\linewidth}\raggedright
Criteria
\end{minipage} & \begin{minipage}[b]{\linewidth}\raggedright
Description
\end{minipage} & \begin{minipage}[b]{\linewidth}\raggedright
Citations
\end{minipage} \\
\midrule\noalign{}
\endhead
\bottomrule\noalign{}
\endlastfoot
Individual Cases & All cases should be treated the same. &
\begin{minipage}[t]{\linewidth}\raggedright
Symmetry requirement\\
\citep{Bourguignon:1979vj}\strut
\end{minipage} \\
Population Size & Proportionate increases or decreases in the size of
the population have no effect on inequality. &
\begin{minipage}[t]{\linewidth}\raggedright
Symmetry axiom for population\\
\citep{Bourguignon:1979vj, Sen:1973vr}\strut \\
Size invariance\\
\citep{James:1985ti, Reardon:2002vt}\strut \\
Population density invariance\\
\citep{Reardon:2004vl}\strut
\end{minipage} \\
Aggregations of Cases & Inequality should be invariant to the
aggregation of components with identical compositions into a single
unit, or dividing a single unit into components with the same
composition. & \begin{minipage}[t]{\linewidth}\raggedright
Organizational equivalence\\
\citep{James:1985ti, Reardon:2002vt}\strut \\
Location equivalence\\
\citep{Reardon:2004vl}\strut \\
Arbitrary boundary independence\\
\citep{Reardon:2004vl}\strut
\end{minipage} \\
\end{longtable}

Measures should satisfy the following criteria about changes to the
whole distribution of cases (Table \ref{tab:criteria2}):

\begin{longtable}[]{@{}
  >{\raggedright\arraybackslash}p{(\columnwidth - 4\tabcolsep) * \real{0.1628}}
  >{\raggedright\arraybackslash}p{(\columnwidth - 4\tabcolsep) * \real{0.4186}}
  >{\raggedright\arraybackslash}p{(\columnwidth - 4\tabcolsep) * \real{0.4186}}@{}}
\caption{Criteria Concerning Changes to the Whole Distribution
\label{tab:criteria2}}\tabularnewline
\toprule\noalign{}
\begin{minipage}[b]{\linewidth}\raggedright
Criteria
\end{minipage} & \begin{minipage}[b]{\linewidth}\raggedright
Description
\end{minipage} & \begin{minipage}[b]{\linewidth}\raggedright
Citations
\end{minipage} \\
\midrule\noalign{}
\endfirsthead
\toprule\noalign{}
\begin{minipage}[b]{\linewidth}\raggedright
Criteria
\end{minipage} & \begin{minipage}[b]{\linewidth}\raggedright
Description
\end{minipage} & \begin{minipage}[b]{\linewidth}\raggedright
Citations
\end{minipage} \\
\midrule\noalign{}
\endhead
\bottomrule\noalign{}
\endlastfoot
Additive Increases & Additive increases to the whole distribution should
reduce inequality, because it reduces the relative difference between
cases. & \begin{minipage}[t]{\linewidth}\raggedright
Scale invariance\\
\citep{Allison:1978vw}\strut
\end{minipage} \\
Proportionate Increases & Multiplying the whole distribution by a
constant should have no effect on inequality, because it has no effect
on the relative difference between cases. &
\begin{minipage}[t]{\linewidth}\raggedright
Scale invariance\\
\citep{Allison:1978vw}\strut \\
Income-zero-homogeneity property\\
\citep{Bourguignon:1979vj}\strut \\
Composition invariance\\
\citep{James:1985ti, Jahn:1947vj, Taeuber:1965us, Morgan:1981to}\strut
\end{minipage} \\
\end{longtable}

The proportionate increases criterion is also known as composition
invariance in the segregation literature. Recent scholarship
\citep[e.g.,][]{Elbers2023ms, Mora:2011ii, Mora2009ip} has renewed
interest in a long debate over whether segregation indexes should be
compositionally invariant, or free from margin dependence.
\citet{James:1985ti} explain the principle of composition invariance
with reference to racial segregation in schools: ``proportional changes
in the numbers of students of a specific race enrolled in each school do
not affect the measured level of segregation'' (p.~16). By their
definition, a segregation index is not compositionally invariant if its
value is a function of the overall population composition, which is also
called margin dependence.

\citet{Coleman:1982vl} argue that under certain definitions of
segregation it is substantively appropriate for an index to be sensitive
to the overall population composition. One such example is defining a
segregation index in terms of the extent of inter-group contact --- no
inter-group contact indicates maximum segregation, and contact
proportional to the overall group proportions indicates zero
segregation. In a population with a small minority group, we could
expect less inter-group contact than in a population with equally
represented groups, and the index adjusts to these expectations. Making
such an index invariant to the population composition would distort its
substantive meaning.

Others take a moderate stance \citep{Reardon:2004vl, Fossett2017nm}, for
example noting that ``the traditional composition invariance criterion
espoused by \citet{James:1985ti} is less important than is ensuring that
a measure of segregation has a sound conceptual basis. If a segregation
index measures exactly that quantity that we believe defines spatial
segregation, then the index will be composition invariant by
definition'' \citep[134]{Reardon:2004vl}.

In recent years, methods have been developed to isolate differences in
segregation that are due to differences in the distribution of the
population across units (e.g., census tract population counts), the
distribution of the population across groups (e.g., racial composition
of a city), and ``structural change''
\citep[e.g.,][]{Elbers2023ms, Mora:2011ii, Mora2009ip}. These three
components are also referred to as unit marginals, group marginals, and
``pure segregation.'' Decomposition analyses can be used to compare how
much each component contributes to overall segregation, and
compositionally invariant versions of indexes have been proposed that
remove the influence of the unit and group marginals.

A key limitation of compositionally invariant indexes is the assumption
that differences in the marginal distributions and differences in the
``structural'' component are independent of one another. There is a
wealth of segregation literature that suggests that changes in unit
marginals and especially group marginals may activate mechanisms
associated with segregation, such as, racially restrictive covenants
\citep[e.g.,][]{Rothstein2017cl}, exclusionary zoning
\citep[e.g.,][]{Rothwell2009ed}, White flight
\citep[e.g.,][]{Lichter:2015gz}, and racial steering by real estate
agents \citep[e.g.,][]{Besbris:2017hn}. A poignant example is the Great
Migration, in which millions of Black people moved from the South to
Northern, Midwestern, and Western states between the 1910s and 1970s.
Racist policies and practices in response to these demographic changes
fueled housing discrimination against Black southerners and created
segregated neighborhoods.

\citet{Elbers2023ms} argues that segregation indexes should be purged of
the influence of marginal differences and only represent the third
component -- ``structural'' (``pure'') segregation. \citet{Mora2009ip}
argue that it is reasonable to work with segregation indexes that
include all three components, and in stylized examples they show that
compositionally invariant indexes can mask important changes in
segregation.

The Divergence Index is not compositionally invariant by design. The
index uses the overall population composition as the comparative
reference for segregation, which captures our expectation about what the
local compositions would be if there is no segregation. Compositional
changes from demographic processes, including migration in and out of
the region and residential resorting within the region, occur and are
relevant for understanding segregation. If there is an influx of a
population group to a city from one time period to the next, the
comparative reference adapts to represent our updated expectation about
the local compositions: the compositions of all local areas should
reflect this influx if there is no segregation in the city.

Measures should satisfy the following criteria about changes within the
distribution (Table \ref{tab:criteria3}):

\begin{longtable}[]{@{}
  >{\raggedright\arraybackslash}p{(\columnwidth - 4\tabcolsep) * \real{0.1591}}
  >{\raggedright\arraybackslash}p{(\columnwidth - 4\tabcolsep) * \real{0.4205}}
  >{\raggedright\arraybackslash}p{(\columnwidth - 4\tabcolsep) * \real{0.4205}}@{}}
\caption{Criteria Concerning Changes within the Distribution
\label{tab:criteria3}}\tabularnewline
\toprule\noalign{}
\begin{minipage}[b]{\linewidth}\raggedright
Criteria
\end{minipage} & \begin{minipage}[b]{\linewidth}\raggedright
Description
\end{minipage} & \begin{minipage}[b]{\linewidth}\raggedright
Citations
\end{minipage} \\
\midrule\noalign{}
\endfirsthead
\toprule\noalign{}
\begin{minipage}[b]{\linewidth}\raggedright
Criteria
\end{minipage} & \begin{minipage}[b]{\linewidth}\raggedright
Description
\end{minipage} & \begin{minipage}[b]{\linewidth}\raggedright
Citations
\end{minipage} \\
\midrule\noalign{}
\endhead
\bottomrule\noalign{}
\endlastfoot
Transfers and Exchanges & \begin{minipage}[t]{\linewidth}\raggedright
1. Any transfer from a unit (e.g., individual, group, or location) with
more of the relevant quantity (e.g., income) to another with less should
decrease inequality, provided that the rank order remains the same.\\
2. Likewise, any transfer to a unit with more of the relevant quantity
should increase inequality.\footnote{For example, from Allison
  \citeyearpar{Allison:1978vw}: ``measures of inequality ought to
  increase whenever income is transferred from a poorer person to a
  richer person, regardless of how poor or rich or the amount of income
  transferred'' (p.~868).}\strut
\end{minipage} & \begin{minipage}[t]{\linewidth}\raggedright
Pigou-Dalton principle\\
\citep{Pigou:1912ux, Dalton:1920ts}\strut \\
Inter-group transfers\\
\citep{James:1985ti, Reardon:2002vt, Reardon:2004vl}\strut \\
Inter-group exchanges\\
\citep{Reardon:2002vt, Reardon:2004vl}\strut
\end{minipage} \\
\end{longtable}

\hypertarget{technical-qualities-and-quantities-of-measures.}{%
\subsubsection{Technical Qualities and Quantities of
Measures.}\label{technical-qualities-and-quantities-of-measures.}}

In addition to desirable conceptual and methodological qualities of
measures, a second set of criteria concern the technical qualities and
quantities of inequality measures. The criteria --- additive
decomposability, and upper and lower bounds are summarized in Table
\ref{tab:criteria4}.

Additive decomposability is a desirable property because it allows for a
deeper analysis of the sources of inequality. The relative contribution
of each component or group to overall inequality can be identified, and
the inequality occurring within- and between-subpopulations can be
analyzed \citep{Bourguignon:1979vj}.

Many measures are bounded between 0 and 1, with 1 indicating maximum
inequality. If a measure has known upper and lower bounds, it can be
rescaled to conform to a 0 to 1 range. However, rescaling the measure
may shift the definition of inequality from absolute to relative. It is
most important for the bounds of the index be known and interpretable.

\begin{longtable}[]{@{}
  >{\raggedright\arraybackslash}p{(\columnwidth - 4\tabcolsep) * \real{0.2088}}
  >{\raggedright\arraybackslash}p{(\columnwidth - 4\tabcolsep) * \real{0.3956}}
  >{\raggedright\arraybackslash}p{(\columnwidth - 4\tabcolsep) * \real{0.3956}}@{}}
\caption{Technical Qualities and Quantities of Measures
\label{tab:criteria4}}\tabularnewline
\toprule\noalign{}
\begin{minipage}[b]{\linewidth}\raggedright
Criteria
\end{minipage} & \begin{minipage}[b]{\linewidth}\raggedright
Description
\end{minipage} & \begin{minipage}[b]{\linewidth}\raggedright
Citations
\end{minipage} \\
\midrule\noalign{}
\endfirsthead
\toprule\noalign{}
\begin{minipage}[b]{\linewidth}\raggedright
Criteria
\end{minipage} & \begin{minipage}[b]{\linewidth}\raggedright
Description
\end{minipage} & \begin{minipage}[b]{\linewidth}\raggedright
Citations
\end{minipage} \\
\midrule\noalign{}
\endhead
\bottomrule\noalign{}
\endlastfoot
Additive Decomposability & Measures should be decomposable into the sum
of inequality within and between sub-populations &
\begin{minipage}[t]{\linewidth}\raggedright
Aggregativity and additivity\\
\citep{Bourguignon:1979vj}\strut \\
Decomposition \citep{Allison:1978vw}\\
Additive decomposability\footnote{For segregation measures, this
  includes additive organizational decomposability
  \citep{Reardon:2002vt}, additive grouping decomposability
  \citep{Reardon:2002vt, Reardon:2004vl} and additive spatial
  decomposability \citep{Reardon:2004vl}.}\\
\citep{Reardon:2002vt, Reardon:2004vl}\strut
\end{minipage} \\
Upper and Lower Bounds & A measure should have known upper and lower
bounds and each should have a substantive interpretation. &
\begin{minipage}[t]{\linewidth}\raggedright
Scale interpretability\\
\citep{Reardon:2004vl}\strut \\
Upper and lower bounds\\
\citep{Allison:1978vw}\strut \\
Principle of Directionality\\
\citep{Fossett:1983jj}\strut
\end{minipage} \\
Relative or Absolute Inequality & Relative and absolute measures are
differentiated based on whether inequality is independent of, or a
function of, the number of categories (respectively). &
\begin{minipage}[t]{\linewidth}\raggedright
Sensitivity to the number of components\\
\citep{Waldman:1977uq}\strut
\end{minipage} \\
\end{longtable}

\hypertarget{summary-of-the-desirable-properties-of-measures}{%
\subsection{Summary of the Desirable Properties of
Measures}\label{summary-of-the-desirable-properties-of-measures}}

Table \ref{tab:evalTable} summarizes the desirable properties of the
Dissimilarity Index, Theil's Inequality Index, the Information Theory
Index, and the Divergence Index. The rows of the table correspond to the
properties detailed in the previous section, as well as the comparative
standard used by the measure and which types of distributions it can be
used with.

\begin{longtable}[]{@{}
  >{\raggedright\arraybackslash}p{(\columnwidth - 8\tabcolsep) * \real{0.1667}}
  >{\centering\arraybackslash}p{(\columnwidth - 8\tabcolsep) * \real{0.2157}}
  >{\centering\arraybackslash}p{(\columnwidth - 8\tabcolsep) * \real{0.2059}}
  >{\centering\arraybackslash}p{(\columnwidth - 8\tabcolsep) * \real{0.2059}}
  >{\centering\arraybackslash}p{(\columnwidth - 8\tabcolsep) * \real{0.2059}}@{}}
\caption{Properties of the Measures
\label{tab:evalTable}}\tabularnewline
\toprule\noalign{}
\begin{minipage}[b]{\linewidth}\raggedright
Criteria
\end{minipage} & \begin{minipage}[b]{\linewidth}\centering
Dissimilarity Index
\end{minipage} & \begin{minipage}[b]{\linewidth}\centering
Theil Index
\end{minipage} & \begin{minipage}[b]{\linewidth}\centering
Information Theory Index
\end{minipage} & \begin{minipage}[b]{\linewidth}\centering
Divergence Index
\end{minipage} \\
\midrule\noalign{}
\endfirsthead
\toprule\noalign{}
\begin{minipage}[b]{\linewidth}\raggedright
Criteria
\end{minipage} & \begin{minipage}[b]{\linewidth}\centering
Dissimilarity Index
\end{minipage} & \begin{minipage}[b]{\linewidth}\centering
Theil Index
\end{minipage} & \begin{minipage}[b]{\linewidth}\centering
Information Theory Index
\end{minipage} & \begin{minipage}[b]{\linewidth}\centering
Divergence Index
\end{minipage} \\
\midrule\noalign{}
\endhead
\bottomrule\noalign{}
\endlastfoot
Individual Cases & \checkmark & \checkmark & \checkmark & \checkmark \\
Population Size & \checkmark & \checkmark & \checkmark & \checkmark \\
Aggregations of Cases & \checkmark & \checkmark & \checkmark &
\checkmark \\
\begin{minipage}[t]{\linewidth}\raggedright
Proportionate\\
Increases\strut
\end{minipage} & X \footnote{It is debatable whether or not the
  Dissimilarity Index satisfies the proportionate increases criterion.
  \citet{Cortese:1976fl} found that it is sensitive to the minority
  group proportion, while others found no such association
  \citep{Taeuber:1965us, James:1985ti, Lieberson:1982vy}. Reardon and
  colleagues \citep{Reardon:2002vt, Reardon:2004vl} find that it is only
  composition invariant when calculated for two groups.} & \checkmark &
X & \checkmark \\
\begin{minipage}[t]{\linewidth}\raggedright
Additive\\
Increases\strut
\end{minipage} & \checkmark & \checkmark & \checkmark & \checkmark \\
Transfers and Exchanges & X \footnote{The Dissimilarity Index satisfies
  a weak form of the transfers and exchanges criteria
  \citep{Reardon:2002vt, Reardon:2004vl}.} & \checkmark & (\checkmark)
\footnote{The transfers and exchanges criterion generally only applies
  when components are mutually exclusive, as described in the text.} &
(\checkmark) \footnotemark
[\value{footnote}] \\
\begin{minipage}[t]{\linewidth}\raggedright
Additive\\
Decomposability\strut
\end{minipage} & X & \checkmark & \checkmark & \checkmark \\
Upper and Lower Bounds & (\checkmark) \footnote{The the Dissimilarity
  Index is bounded between 0 and 1, but the expected value of the index
  is greater than 0 \citep{Cortese:1976fl, Mazza2015ub}.} & \checkmark &
\checkmark & \checkmark \\
Relative or Absolute Inequality & Relative & Either & Absolute &
Either \\
\begin{minipage}[t]{\linewidth}\raggedright
Comparative\\
Standard\strut
\end{minipage} & \begin{minipage}[t]{\linewidth}\centering
Evenness\\
(mean of the distribution)\strut
\end{minipage} & \begin{minipage}[t]{\linewidth}\centering
Evenness\\
(mean of the distribution)\strut
\end{minipage} & Randomness & Any \\
\begin{minipage}[t]{\linewidth}\raggedright
Distribution\\
Types\strut
\end{minipage} & \begin{minipage}[t]{\linewidth}\centering
Discrete with\\
nominal categories\strut
\end{minipage} & Continuous & Discrete &
\begin{minipage}[t]{\linewidth}\centering
Discrete or\\
continuous\strut
\end{minipage} \\
Citations & \citep[
\citet{Mazza2015ub}]{Bourguignon:1979vj, Cortese:1976fl, Coulter:1989us, Duncan:1955ve, Falk:1978uz, Fossett:1983jj, Jahn:1947vj, James:1985ti, Lieberson:1982vy, Massey:1988tq, Morgan:1975vu, Reardon:2002vt, Reardon:2004vl, Sakoda:1981uw, Taeuber:1965us, Theil:1972vb, Winship:1978tm}
&
\citep{Allison:1978vw, Bourguignon:1979vj, Theil:1967vj, Cowell:1980un, Shorrocks:1980hi, Shorrocks:1984ip, Shorrocks:2012cv, Cowell:2013ir, Theil:1972vb}
&
\citep{Reardon:2002vt, Reardon:2004vl, Theil:1967vj, Theil:1972vb, White:1986vx}
&
\citep{Bavaud:2009fn, Cover:2006ub, Theil:1967vj, Magdalou:2011ci, Mori:2005wu, Shorrocks:1980hi, Shorrocks:1984ip, Shorrocks:2012cv, Cowell:1980un, Cowell:2013ir, Walsh:1979um} \\
\end{longtable}

The Information Theory Index does not satisfy the proportionate
increases criterion according to the definition of composition
invariance described by \citet{James:1985ti} --- the value of the index
should not be a function of the overall population composition. However,
\citet{Reardon:2004vl} show that the index does conform to other
definitions of composition invariance. For instance, it is invariant to
compositional changes as long as the relationship between local
population diversity and overall population diversity remains constant.

\citet{Reardon:2004vl} show that the Information Theory Index satisfies
the transfers and exchanges criteria when used to measure aspatial
segregation. None of the indexes they evaluated satisfy the transfers
criterion when used to measure spatial segregation. Spatial approaches
often include a proximity weighted contribution from neighboring areas
in each location's population. This makes it difficult for any index to
satisfy the transfers and exchanges criteria because the local
populations are not mutually exclusive. They show that the Information
Theory Index satisfies the exchanges criterion under certain general
conditions \citep[see][]{Reardon:2004vl}.

The entropy-based measures (Theil Index, Information Theory Index, and
Divergence Index) can be defined using logarithms to any base. The
selected base defines the units of the index
\citep{Shannon:1948iy, Theil:1972vb}. Log base 2 (\(\log_2\)) is
typically used in information theory, which gives results in units of
binary bits of information. It is common for inequality measures to use
the natural logarithm (\(\ln\)), which has the mathematical constant
(\(e\)) as its base.

Using a fixed log base, such as base 2 (\(\log_2\)) or \(e\) (\(\ln\)),
entropy is an absolute measure. Results are a function of the number of
groups in the population \citep{Waldman:1977uq}. Given a uniform
distribution of groups (indicating maximum diversity), entropy is an
increasing function of the number of groups. At first blush, this may
seem undesirable, but it has the benefit of maintaining entropy's
aggregation equivalence and independence. This means that inequality
calculated for a population of two groups is the same as if there were
three groups in same population, but no individuals associated with the
third type.

For discrete distributions, it may be preferable to use the number of
groups as the base. The result is equivalent to dividing by the maximum
entropy (\(\log{M}\)), given by the number of groups (\(M\)). With the
number of groups as the log base (\(\log_M\)), values are scaled to have
the same maximum entropy no matter how many groups are in the
population. This transforms entropy from an absolute to a relative
measure of inequality. It allows for easier comparison across indexes
with different numbers of groups, but comes at the cost of one of the
desirable properties of entropy --- aggregation equivalence and
independence.

For example, using \(\log_2\) to measure Black-Latinx-White residential
segregation in a city with no Latinx residents gives the same values
whether all three races are included in the measure or only the two with
population. This is not the case using \(\log_M\), because the values
are scaled according to the number of groups included in the index.
Which of these options is preferable depends on the analytic aim of the
research, but it is important to be aware of this trade-off.\footnote{This
  choice does not affect the values of the information theory index,
  because the log appears both in the numerator and denominator of the
  equation.}

\clearpage

\setcounter{table}{0}
\renewcommand{\thetable}{C\arabic{table}}

\hypertarget{appendix-c}{%
\section{Appendix C}\label{appendix-c}}

\hypertarget{entropy-decomposition}{%
\subsection{Entropy Decomposition}\label{entropy-decomposition}}

Entropy-based measures are additively decomposable, which is a
particularly desirable property \citep{Theil:1972vb}.\footnote{The
  additivity of entropy comes from one of the properties of logarithms:
  \(log{\left(\pi_1\cdot\pi_2\right)}=log{\left(\pi_1\right)}+log{\left(\pi_2\right)}\)}
It is simple to aggregate (and disaggregate) the entropy for multiple
groups and to decompose total entropy into the entropy occurring within-
and between-groups. The entropy for each component (\(i\)) is the sum of
the entropy across groups within that component (\(m\)):
\[E_i=\sum_{m=1}^{M}{\pi_{im}\log{\cfrac{1}{\pi_{im}}}}\] The entropy
for all components is the mean of the individual entropies, weighted by
the relative size of each component:
\[\bar{E_i}=\sum_{i=1}^{N}{\cfrac{\tau_i}{T}E_i}\]

\citet{Theil:1972vb} showed that total entropy can be calculated for any
subdivision of the population and written as the sum of a
between-subdivision entropy and the average within-subdivision
entropies. For example, if the groups are aggregated into supergroups
(\(S_g\)), where \(\Pi_{ig}=\sum_{m\in S_g}{\pi_{im}}\) is the
proportion in each supergroup (\(g\)) within component (\(i\)). The
entropy within supergroup \(g\) for component \(i\) is:
\[E_{ig}=\sum_{m\in S_g}{\cfrac{\pi_{im}}{\Pi_{ig}}\log{\cfrac{\Pi_{ig}}{\pi_{im}}}}\]
And the between-supergroup entropy is:
\[E_{i0}=\sum_{g=1}^{G}{\cfrac{\Pi_{ig}}{\pi_{i.}}\log{\cfrac{\pi_{i.}}{\Pi_{ig}}}}\]
The total entropy for component \(i\) can then be written as the
between-supergroup entropy (\(E_{i0}\)) plus the average
within-supergroup entropy
(\(E_{ig}\)):\[E_i=E_{i0}+\sum_{g=1}^{G}{\cfrac{\Pi_{ig}}{\pi_{i.}}E_{ig}}\]

\clearpage

\setcounter{table}{0}
\renewcommand{\thetable}{D\arabic{table}}

\hypertarget{appendix-d}{%
\section{Appendix D}\label{appendix-d}}

\hypertarget{comparing-the-theil-index-and-divergence-index}{%
\subsection{Comparing the Theil Index and Divergence
Index}\label{comparing-the-theil-index-and-divergence-index}}

Theil's inequality index (\(I\)) and the Divergence Index (\(D\)) both
measure inequality relative to a defined standard. The Theil Index
measures the difference between the observed shares of income across
individuals or groups and a theoretical uniform distribution --- one in
which everyone's income is equal to the mean.

There is a straightforward equivalency between \(I\) and \(D\) for
continuous distributions, such as income.\footnote{Moreover, the
  equivalency applies to any distribution for which a mean can be
  calculated, such as a discrete simplification of a continuous
  distribution.} Theil's index can be written like the Divergence Index,
where \(P_i\) is \(i\)'s share of total aggregate income,
\(\cfrac{\tau_i x_i}{T\bar{x}}\) , and \(Q_i\) is the theoretical
uniform share \(\cfrac{\tau_i}{T}\) : \begin{align*}
  D\left(P\parallel Q\right) &= \sum_{m=1}^{M}{P_m\log{\cfrac{P_m}{Q_m}}} \\
  I &=\sum_{i=1}^{N}{P_i\log{\cfrac{P_i}{Q_i}}} \\
  &=\sum_{i=1}^{N}{\cfrac{\tau_ix_i}{T\bar{x}}\log{\cfrac{\cfrac{\tau_ix_i}{T\bar{x}}}{\cfrac{\tau_i}{T}}}} \\
  &=\cfrac{1}{T}\sum_{i=1}^{N}{\cfrac{\tau_ix_i}{\bar{x}}\log{\cfrac{x_i}{\bar{x}}}}
\end{align*} If \(\tau_i=1\) and \(T=N\), then we get:
\[I=\cfrac{1}{N}\sum_{i=1}^{N}{\cfrac{x_i}{\bar{x}}\log{\cfrac{x_i}{\bar{x}}}}\]
We can see that \(I\) is a specific case of \(D\) applied to measuring
income inequality, using uniform shares of income as the comparative
standard.

\clearpage

\setcounter{figure}{0}
\renewcommand{\thefigure}{E\arabic{figure}}

\setcounter{table}{0}
\renewcommand{\thetable}{E\arabic{table}}

\hypertarget{appendix-e}{%
\section{Appendix E}\label{appendix-e}}

\begin{table}[h]
  \captionsetup{width=\linewidth}
  \caption{Asian-Black-Latinx-White Entropy Index from 1990 to 2010, Mean for the 100 Largest Metropolitan Areas
  \label{tab:metroE}}
  \includegraphics[width=4in, center]{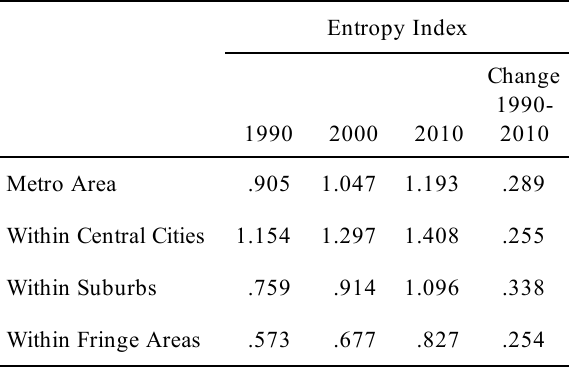}
  \begin{minipage}{\linewidth}
  \vspace{2ex}
  \small Data Source: Author's calculations of block-level population data from the decennial census using 2010 boundaries \citep{CensusSummary:bZv73ozJ, NHGIS:2021}
  \end{minipage}
\end{table}

\begin{longtable}[]{@{}
  >{\raggedright\arraybackslash}p{(\columnwidth - 6\tabcolsep) * \real{0.1944}}
  >{\centering\arraybackslash}p{(\columnwidth - 6\tabcolsep) * \real{0.2361}}
  >{\centering\arraybackslash}p{(\columnwidth - 6\tabcolsep) * \real{0.2361}}
  >{\centering\arraybackslash}p{(\columnwidth - 6\tabcolsep) * \real{0.1111}}@{}}
\caption{Number of Metros with Changes in Within-City and Within-Suburb
Index Values for the Divergence Index in the 100 Largest Metropolitan
Areas, 1990 to 2010 \label{tab:withinDcount}}\tabularnewline
\toprule\noalign{}
\begin{minipage}[b]{\linewidth}\raggedright
~
\end{minipage} & \begin{minipage}[b]{\linewidth}\centering
Within-Suburbs Decrease
\end{minipage} & \begin{minipage}[b]{\linewidth}\centering
Within-Suburbs Increase
\end{minipage} & \begin{minipage}[b]{\linewidth}\centering
Total
\end{minipage} \\
\midrule\noalign{}
\endfirsthead
\toprule\noalign{}
\begin{minipage}[b]{\linewidth}\raggedright
~
\end{minipage} & \begin{minipage}[b]{\linewidth}\centering
Within-Suburbs Decrease
\end{minipage} & \begin{minipage}[b]{\linewidth}\centering
Within-Suburbs Increase
\end{minipage} & \begin{minipage}[b]{\linewidth}\centering
Total
\end{minipage} \\
\midrule\noalign{}
\endhead
\bottomrule\noalign{}
\endlastfoot
Within-City Decrease & 12 & 46 & 58 \\
Within-City Increase & 4 & 38 & 42 \\
Total & 16 & 84 & 100 \\
\end{longtable}

\begin{minipage}{\linewidth}
  \vspace{-10ex}
  \singlespacing \small Data Source: Author's calculations of block-level population data from the decennial census using 2010 boundaries \citep{CensusSummary:bZv73ozJ, NHGIS:2021}
  \end{minipage}

\begin{longtable}[]{@{}
  >{\raggedright\arraybackslash}p{(\columnwidth - 6\tabcolsep) * \real{0.1944}}
  >{\centering\arraybackslash}p{(\columnwidth - 6\tabcolsep) * \real{0.2361}}
  >{\centering\arraybackslash}p{(\columnwidth - 6\tabcolsep) * \real{0.2361}}
  >{\centering\arraybackslash}p{(\columnwidth - 6\tabcolsep) * \real{0.1111}}@{}}
\caption{Number of Metros with Changes in Within-City and Within-Suburb
Index Values for the Information Theory Index in the 100 Largest
Metropolitan Areas, 1990 to 2010
\label{tab:withinHcount}}\tabularnewline
\toprule\noalign{}
\begin{minipage}[b]{\linewidth}\raggedright
~
\end{minipage} & \begin{minipage}[b]{\linewidth}\centering
Within-Suburbs Decrease
\end{minipage} & \begin{minipage}[b]{\linewidth}\centering
Within-Suburbs Increase
\end{minipage} & \begin{minipage}[b]{\linewidth}\centering
Total
\end{minipage} \\
\midrule\noalign{}
\endfirsthead
\toprule\noalign{}
\begin{minipage}[b]{\linewidth}\raggedright
~
\end{minipage} & \begin{minipage}[b]{\linewidth}\centering
Within-Suburbs Decrease
\end{minipage} & \begin{minipage}[b]{\linewidth}\centering
Within-Suburbs Increase
\end{minipage} & \begin{minipage}[b]{\linewidth}\centering
Total
\end{minipage} \\
\midrule\noalign{}
\endhead
\bottomrule\noalign{}
\endlastfoot
Within-City Decrease & 78 & 16 & 94 \\
Within-City Increase & 3 & 3 & 6 \\
Total & 81 & 19 & 100 \\
\end{longtable}

\begin{minipage}{\linewidth}
  \vspace{-10ex}
  \singlespacing \small Data Source: Author's calculations of block-level population data from the decennial census using 2010 boundaries \citep{CensusSummary:bZv73ozJ, NHGIS:2021}
  \end{minipage}

\clearpage
\pagenumbering{arabic}
\setcounter{page}{0}

\end{document}